\documentclass[%
 preprint,
 nofootinbib,
 amsmath,amssymb,
 aps,
 prd,
 numbers,sort&compress,
 longbibliography
]{revtex4-2}

\usepackage{graphicx}  
\usepackage{float}
\usepackage{dcolumn}   
\usepackage{bm}        
\usepackage{amssymb}   
\usepackage{slashed}   
\usepackage{amsmath}   
\usepackage{simplewick} 
\usepackage{verbatim}  
\usepackage{color} 
\usepackage{xcolor} 
\usepackage{appendix} 
\usepackage{subfig} 
\usepackage{epstopdf}

\definecolor{Blue}{rgb}{0, 0.1, 0.5}

\usepackage[bookmarks=true,colorlinks=true,linkcolor=blue,unicode=true]{hyperref}

\newcounter{YJC}

\usepackage{multirow}
\begin{document}

\title{Searching for gluon quartic gauge couplings at muon colliders using the auto-encoder}

\author{Yu-Ting Zhang}
\email{976982345@qq.com}
\author{Xin-Tong Wang}
\email{2226769965@qq.com}

\affiliation{Department of Physics, Liaoning Normal University, Dalian 116029, China}

\author{Ji-Chong Yang}
\email{yangjichong@lnnu.edu.cn}
\thanks{Corresponding author}

\affiliation{Department of Physics, Liaoning Normal University, Dalian 116029, China}
\affiliation{Center for Theoretical and Experimental High Energy Physics, Liaoning Normal University, Dalian 116029, China}

\begin{abstract}
One of the difficulties one has to face in the future phenomenological studies of the new physics~(NP), is the need to deal with increasing amounts of data. 
It is therefore increasingly important to improve the efficiency in the phenomenological study of the NP. 
Whether it is the use of the Standard Model effective field theory~(SMEFT), the use of machine learning~(ML) algorithms, or the use of quantum computing, all are means of improving the efficiency. 
In this paper, we use a ML algorithm, the auto-encoder~(AE), to study the dimension-8 operators in the SMEFT which contribute to the gluon quartic gauge couplings~(gQGCs) at muon colliders. 
The AE is one of the ML algorithms that has the potential to be accelerated by the quantum computing.
It is found that the AE-based anomaly detection algorithm can be used as event selection strategy to study the gQGCs at the muon colliders, and is effective compared with traditional event selection strategies.
\end{abstract}

\maketitle

\section{\label{sec:1}Introduction}

Supported by the large amount of experimental evidences, it can be concluded that the Standard Model~(SM) is able to describe and explain the vast majority of phenomena in particle physics, with a few rare exceptions.
These exceptions include experimental results such as the neutrino mass~\cite{Farzan:2017xzy,Proceedings:2019qno,Arguelles:2022tki}, the $W$ boson mass problem~\cite{CDF:2022hxs,deBlas:2022hdk}, the muon $g-2$ problem~\cite{Muong-2:2021ojo,Muong-2:2023cdq,Aoyama:2020ynm}, and more~\cite{Crivellin:2023zui}.
Besides, the SM cannot describe dark matter, gravity, etc.
As a result, the existence of the new physics~(NP) beyond the SM has been widely believed, and the search of NP as well as precision measurements have been at the forefront of interest in the high energy physics~(HEP) community~\cite{Ellis:2012zz}.

Both the search for NP and precision measurements require dealing with a large number of events.
With more and more data to be processed, more efficient ways to search for NP are called for.
One of the reasons why the SM effective field theory~(SMEFT)~\cite{Weinberg:1979sa,Grzadkowski:2010es,Willenbrock:2014bja,Masso:2014xra} has been widely used in the phenomenological study of NP in recent years is that SMEFT has an outstanding advantage of searching for NP signals with high efficiency.
In the SMEFT, the NP particles are integrated out, and the NP effects become new interactions of known particles. 
Formally, the new interactions appear as higher dimension operators with Wilson coefficients suppressed by powers of a NP scale $\Lambda$. 
The operators that are most likely to be found correspond to the operators with Wilson coefficients that are least suppressed by $\Lambda$.
The high efficiency of searching for NP using SMEFT is demonstrated by the fact that, instead of dealing with various NP models, the number of operators to be considered at a specific order of $\Lambda$ in the SMEFT is finite.
Not only that, but if an operator is not found, then multiple NP models contributing to this operator will also be constrained.
However, as the importance of the dimension-8 operators in the SMEFT has been realized~\cite{Green:2016trm,Zhang:2020jyn,Murphy:2020rsh,Li:2020gnx}, more and more phenomenological studies have been devoted to the dimension-8 operators in recent years~\cite{Zhang:2018shp,Bi:2019phv,ATLAS:2017vqm,CMS:2017rin,CMS:2020ioi,CMS:2016gct,CMS:2017zmo,CMS:2018ccg,ATLAS:2018mxa,CMS:2019uys,CMS:2016rtz,CMS:2017fhs,Ellis:2023ucy,Spor:2022hhn,Spor:2022zob,Yilmaz:2021ule,Ellis:2020ljj,Senol:2019swu,Yilmaz:2019cue,Ellis:2019zex,Guo:2020lim,Guo:2019agy,Yang:2021pcf,Fu:2021mub,Yang:2020rjt,Jahedi:2022duc,Jahedi:2023myu}.
For one generation of fermions, there are $895$ baryon number conserving dimension-8 operators~\cite{Anders:2018oin,Henning:2015alf}, and there are even more operators if one considers operators with dimension more than eight.
A procedure to select the events which does not rely on operators to be searched for can further improve efficiency.

Machine learning~(ML) algorithms are one of the ways to process the data efficiently.
ML is a multidisciplinary cross-discipline for studying how computers can mimic and implement human learning behaviors in order to acquire new knowledge, and has already been used in HEP studies~\cite{Radovic:2018dip,Baldi:2014kfa,Ren:2017ymm,Abdughani:2018wrw,Ren:2019xhp,Letizia:2022xbe,DAgnolo:2019vbw,DAgnolo:2018cun,Yang:2021ukg,Yang:2022fhw}.
ML algorithms have an advantage in processing complex data, and one of its common applications is anomaly detection~(AD). 
When using AD to search for NP models, its implementation is often independent of the NP model to be searched for. 
While hyperparameters in ML algorithms will often be NP model-dependent, the procedure of tuning of hyperparameters is usually common for different NP models.
Based on this, an event selection strategy utilizing AD can be viewed as an automated event selection strategy.
The use of AD algorithms in phenomenological studies is a hot topic in recent years~\cite{DeSimone:2018efk,MdAli:2020yzb,Fol:2020tva,Kasieczka:2021xcg,Guo:2021jdn,Yang:2021kyy,Dong:2023nir,CrispimRomao:2020ucc,vanBeekveld:2020txa,Kuusela:2011aa}.

Meanwhile, quantum computing is another effective way to deal with large amounts of data. 
Many ML algorithms can be implemented by quantum computing~\cite{Biamonte:2017cqe,qml2,Garcia:2022cqq}, an example of which is the auto-encoder~(AE) algorithm~\cite{LIOU20083150,LIOU201484}.
AE is an unsupervised learning dimensionality reduction algorithm using artificial neural networks~(ANN), which is at the same time capable of AD. 
Therefore, it can be expected that AE can be used to detect NPs, and AE has already been used in phenomenological studies in HEP~\cite{CrispimRomao:2020ucc,vanBeekveld:2020txa,Farina:2018fyg,Cerri:2018anq}.
Analogous to the principal component analysis~(PCA) algorithm, the most common scenario for AE is data dimensionality reduction. 
While the PCA is a linear dimensionality reduction by solving the feature vector, the AE algorithm is a nonlinear dimensionality reduction.
It is verified that, AD based on PCA is able to search for NP~\cite{Dong:2023nir}, therefore it can be expected that a similar approach works also for AE.
Similar to PCA, AE has the potential for quantum acceleration~\cite{Romero_2017,Bravo-Prieto_2021,PhysRevLett.124.130502,qvae}, and thus AE holds great promise for processing large amounts of data.
To verify the feasibility of the AE algorithm in the search of high dimensional operators in the SMEFT, we use an AE algorithm based AD~(AEAD) to study the gluon quartic gauge couplings~(gQGCs)~\cite{Ellis:2018cos,Ellis:2021dfa} in this paper.
In the SMEFT, operators contributing to gQGCs start from dimension-8, therefore we concentrate on those dimension-8 operators.

As an arena, the processes affected by gQGCs at muon colliders are considered. 
The muon colliders have been hotly discussed in recent years for searching for NP signals~\cite{Buttazzo:2018qqp,Delahaye:2019omf,Lu:2020dkx,Franceschini:2021aqd,Palmer:1996gs,Holmes:2012aei,Costantini:2020stv,AlAli:2021let,Han:2020uid,Han:2021kes,Aime:2022flm}. 
A muon collider has the advantage of being able to explore both high luminosity and high energy frontiers, while at the same time being less affected by the QCD background as a lepton collider. 
On the one hand, higher collision energies are better if one is committed to studying dimension-8 or even higher dimension operators, and on the other hand, higher luminosities require more efficient means of processing data. 
Thus, the processes at muon colliders that are affected by gQGCs are both worth studying and suitable for exploring the AEAD algorithm. 
For comparison with a conventional event selection strategy, in this paper the process $\mu^+\mu^- \to j j \nu \bar{\nu}$ is studied which has been studied in Ref.~\cite{Yang:2023gos}. 
It has been shown that this process is sensitive to the gQGCs.

The rest of the paper is organized as follows, in section~\ref{sec:2}, the dimension-8 operators contributing to the gQGCs are introduced. 
In section~\ref{sec:3}, the event selection strategy based on AEAD algorithm is presented.
In section~\ref{sec:4}, the numerical results and expected constraints on the operator coefficients are presented. 
Section~\ref{sec:5} is a summary.

\section{\label{sec:2}Dimension-8 operators contributing to the gQGCs}

\begin{table}[!htbp]
\begin{center}
\begin{tabular}{c|c}
\hline
${M_0}\geq 1040\;{\rm GeV}$ & ${M_1}\geq 777\;{\rm GeV}$ \\
\hline
${M_2}\geq 750\;{\rm GeV}$ & ${M_3}\geq 709\;{\rm GeV}$ \\ 
\hline
${M_4}\geq 1399\;{\rm GeV}$ & ${M_5}\geq 1046\;{\rm GeV}$ \\
\hline
${M_6}\geq 1010\;{\rm GeV}$ & ${M_7}\geq 954\;{\rm GeV}$ \\
\hline
\end{tabular}
\caption{\label{tab.currentconstraint}The constraints on the operator coefficients at 95\% CL obtained by the process $gg\to \gamma\gamma$ at the LHC with $\sqrt{s}=13\;{\rm TeV}$ and $L=36.7\;{\rm fb}^{-1}$~\cite{Ellis:2018cos}.}
\end{center}
\end{table}

\begin{figure}[!htbp]
\centering{
\includegraphics[width=0.98\hsize]{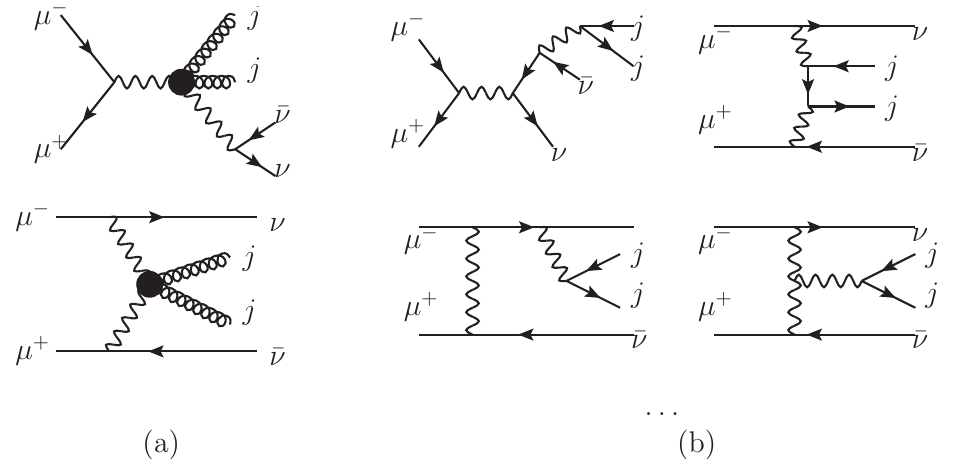}
\caption{\label{fig:feyndiag}Feynman diagrams for the process $\mu^+\mu^-\to j j \nu \bar{\nu}$, where (a) shows the Feynman diagrams of the gQGC contribution, and (b) shows the typical Feynman diagrams of the SM background.}}
\end{figure}

The gQGCs arise from the Born-Infeld~(BI) extension of the SM, which was originally motivated by the idea that there should be an upper limit on the strength of the electromagnetic field~\cite{Born:1934gh}.
It has been shown that, the BI model is also related to the M-theory inspired models~\cite{Ellis:2018cos,Ellis:2021dfa,Fradkin:1985qd,Tseytlin:1999dj,Cheung:2018oki}.
In the SMEFT, the operators contributing to gQGCs appear at dimension-8,
\begin{equation}
\begin{aligned}
O_{gT,0}=\frac{1}{16M_0^4}\sum_a G_{\mu\nu}^a G^{a,\mu\nu}\times\sum_i W_{\alpha\beta}^i W^{i,\alpha\beta},\\
O_{gT,1}=\frac{1}{16M_1^4}\sum_a G_{\alpha\nu}^a G^{a,\mu\beta}\times\sum_i W_{\mu\beta}^i W^{i,\alpha\nu},\\
O_{gT,2}=\frac{1}{16M_2^4}\sum_a G_{\alpha\mu}^a G^{a,\mu\beta}\times\sum_i W_{\nu\beta}^i W^{i,\alpha\nu},\\
O_{gT,3}=\frac{1}{16M_3^4}\sum_a G^a_{\alpha \mu} G^a_{\beta \nu} \times \sum_i W^{i, \mu \beta} W^{i, \nu \alpha},\\
O_{gT,4}=\frac{1}{16M_4^4}\sum_a G^a_{\mu \nu} G^{a, \mu \nu} \times B_{\alpha \beta} B^{\alpha \beta},\\
O_{gT,5}=\frac{1}{16M_5^4}\sum_a G_{\alpha\nu}^a G^{a,\mu\beta}\times B_{\mu\beta}B^{\alpha\nu},\\
O_{gT,6}=\frac{1}{16M_6^4}\sum_a G_{\alpha\mu}^a G^{a,\mu\beta}\times B_{\nu\beta}B^{\alpha\nu},\\
O_{gT,7}=\frac{1}{16M_7^4}\sum_a G^a_{\alpha \mu} G^a_{\beta \nu}\times B^{\mu \beta} B^{\nu \alpha},\\
\end{aligned}
\end{equation}
where ${G_{\mu\nu}^a}$ is gluon field strengths, ${W_{\mu\nu}^i}$ and ${B_{\mu\nu}}$ denote electroweak field strengths, and $M_i$ are mass scales associated with NP particles.
For convenience, we define $f_{i}\equiv 1/(16M_i^4)$.
The expected constraints on $M_i$ at the Large Hadron Collider~(LHC) with the center of mass~(c.m.) energy $\sqrt{s}=13\;{\rm TeV}$, and luminosity $L=36.7\;{\rm fb}^{-1}$ obtained by using the process $gg\to \gamma\gamma$ are shown in Table~\ref{tab.currentconstraint}.
The combined sensitivities of the $Z\gamma$ and $\gamma\gamma$ channels at the LHC with $\sqrt{s}=13\;{\rm TeV}$ and $L=137\;{\rm fb}^{-1}$~\cite{Ellis:2021dfa} are about three times of the ones shown in Table~\ref{tab.currentconstraint}.

The process $\mu^+\mu^-\to j j \nu \bar{\nu}$ at muon colliders can also be affected by the gQGCs~\cite{Yang:2023gos}.
Different from the case of a hadron collider that the operators are classified into four pairs with same Lorentz structures in phenomenological studies, at the muon colliders, the pairs can be decoupled.
In particular, the process $\mu^+\mu^-\to j j \nu \bar{\nu}$ can be affected by the vector boson scattering~(VBS) subprocess $W^+W^-\to gg$, which is associated with only $O_{gT,0,1,2,3}$ operators.
The Feynman diagrams are shown in Fig.~\ref{fig:feyndiag}.
Since the VBS contribution is logarithmically enhanced at large c.m. energies compared with the tri-boson process, we concentrate on the $O_{gT,0,1,2,3}$ operators.
It has been shown that, at muon colliders the process $\mu^+\mu^-\to j j \nu \bar{\nu}$ can archive a competitive sensitivity compared with the hadron colliders~\cite{Yang:2023gos}.

In the process $\mu^+\mu^-\to j j \nu \bar{\nu}$, there is no interference between the SM and gQGCs, which simplifies the procedure to obtain the expected constraints.
However, there are two (anti-)neutrinos in the final state. 
This usually results in some loss of information, which in turn affects the efficiency of the event selection strategy.
This just provides a place to test whether the AEAD algorithm is effective or not.

\section{\label{sec:3}Auto-encoder anomaly detection}

\subsection{\label{sec:3.1}A brief introduction of the auto-encoder algorithm}

\begin{figure}[htbp]
\centering
\includegraphics[width=0.98\hsize]{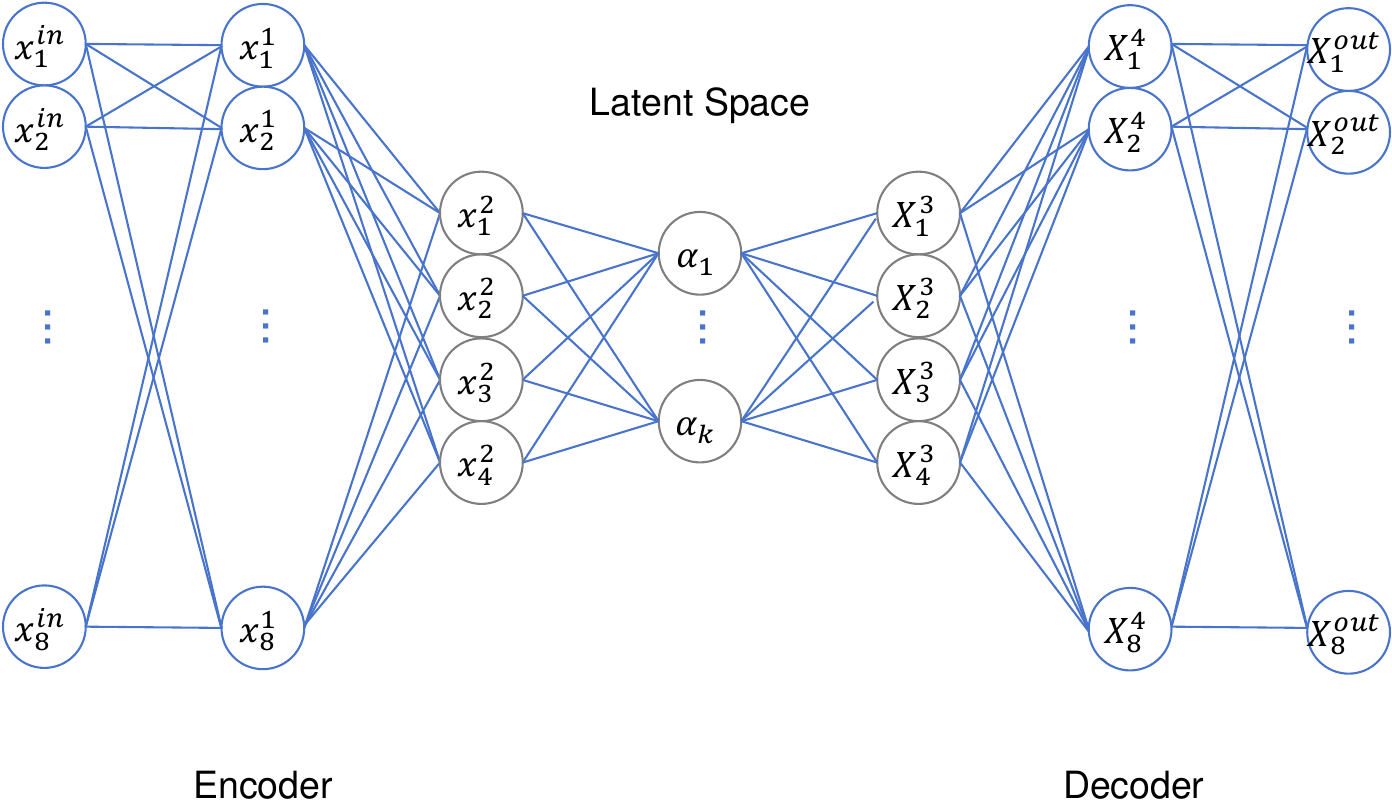}
\caption{\label{fig:aestructure}
The graphical representation of the AE. The AE network can be decomposed into two parts, where the encoder is consist of the $x^{in,1,2}$ and the latent layers, the decoder is consist of the latent, $X^{3,4,out}$ layers. In this paper, we use a dense connected network.}
\end{figure}

AE is a type of ANN, which can be used in various applications, including data compression, feature extraction, denoising, data generation, etc. 
The structure of an AE is shown in Fig.~\ref{fig:aestructure}, which is primarily composed of two parts, the encoder and the decoder. 
Both encoder and decoder can consist of multiple layers, with the neurons in the input layer denoted as ${x_i^{in}}$, and those in the output layer denoted as ${X_i^{out}}$.
The data input to the input neurons and the data obtained from the output neurons are also notated as ${x_{i}^{in}}$ and ${X_{i}^{out}}$, respectively.
The number of neurons in the input layer is as same as the one in the output layer and the dimension of the input vector~(denoted as $l$). 
In this paper, $l=8$.
The goal to train an AE is to reconstruct the input data, i.e., the labels of the training data are just the input data, therefore the training of an AE is unsupervised learning.
The training objective is to minimize the reconstruction error between the ${\vec{x}^{in}}$ and ${\vec{X}^{out}}$, aiming to obtain as similar a set of vectors from the output layer as possible after inputting a set of vectors into the input layer.

A well-trained AE can reproduce the input by utilizing a few variables, $\alpha _{i=1,\ldots k}$ with $k< l$, along with the decoder network. 
Being able to reconstruct the input data indicates that, the information in $\alpha _i$ is enough to describe the data with the help of the decoder.
Also, this means that the encoder is able to compress the information in the input data into $k$ numbers, $\alpha _i$~(which is often called the latent space), i.e., data dimensionality reduction.

The reason an AE can achieve data dimensionality reduction lies in the fact that, the features of the input data are not independent.
The encoder learns the relationships among the features and compresses them into $\alpha_i$ variables, which can then be used by the decoder to reconstruct the ${\vec{X}^{out}}$.
This mechanism of AE can be used for AD.
It is well known that, a challenge for ML methods is the ability to extrapolate into unknown phase space regions, AEAD just turned the disadvantage in extrapolating into a feature, i.e., the artificial intelligence trained using the SM cannot recognize NP and will treat NP as anomalies.

At first, AE is trained on the SM events. 
Since the events are generated according to the physical laws of the SM, the relationships between the features learned by the AE are also related to the physical laws of the SM shaded behind them.
At the same time, it can be expected that the reconstruction of the events generated according to the physical laws of the NP will be less accurate than that of the SM, because the AE has not learned the relationships between the features of the events of the NP.
The mean squared error~(mse) defined as $d=\sum _n^N \left(\vec{x}^{in,n}-\vec{X}^{out,n}\right)^2/N$ can be used to represent the reconstruction error, where $N$ is the number of events.
In AEAD, $d$ can also been used as an anomaly score.
It can be expected that $d$ is larger for the NP events compared with those of the SM events, therefore can be used to select the NP events.

\subsection{\label{sec:3.2}The event selection strategy based on AEAD}

Following the above idea, the AEAD event selection strategy can be summarized as follows,
\begin{itemize}
    \item Generate the training and validation data sets using Monte Carlo~(MC) simulation, which consist of the events from the SM.
    \item Train AE, and use the validation data set to avoid overfitting.
    \item For a test data set, which can be either obtained from the experiment or from MC simulation, calculate $d$ for each event.
    \item Use a threshold $d_{th}$ to select events, i.e. select the events with $d>d_{th}$, where $d_{th}$ can be tuned according to the signal significance or expected constraints on parameters of NP.
\end{itemize}

Note that, although the AE is unsupervised ML algorithm, in AEAD the SM data set is used in the training phase, therefore the AEAD is no longer unsupervised.
However, the NP data sets are not used in the training phase.
The test data set can be from the experiment, and there is no need to know whether it contains NP or what kind of NP it might contain.
It can be expected that, the signal of NP can be traced by the AEAD whenever the test data set contains events that differ from the law of the SM.

\section{\label{sec:4}Numerical results}

\subsection{\label{sec:41}Data preparation}

\begin{table}[htbp]
\centering
\begin{tabular}{c|c|c|c|c}
\hline
 $\sqrt{s}\;({\rm TeV})$  & 3 & 10 & 14 & 30 \\
\hline 
 $\sigma _{\rm SM}\;({\rm fb})$ & $868.8$ &  $1454.8$ &  $1608.7$ &  $1898.8$ \\
 $\hat{\sigma} _{\rm SM}\;({\rm fb})$ & $722.9$ &  $1191.9$ &  $1315.5$ &  $1542.4$ \\
\hline
 $f_0^U\;({\rm TeV}^{-4})$ &  3.5 &  0.028 & 0.0074 & 0.00035 \\
 $\tilde{f}_0\;({\rm TeV}^{-4})$ & 1 & 0.012 &  0.004 &  0.00035 \\
 $\sigma _{gT,0}\left(\tilde{f}_0\right)\;({\rm fb})$ & $3.21$ &  $1.15$ &  $1.12$ &  $1.14$ \\
 $\hat{\sigma} _{gT,0}\left(\tilde{f}_0\right)\;({\rm fb})$ & $3.21$ &  $1.15$ &  $1.12$ &  $1.14$ \\
\hline
 $f_1^U\;({\rm TeV}^{-4})$ &  10.5 &  0.085 & 0.022 &  0.001 \\
 $\tilde{f}_1\;({\rm TeV}^{-4})$ & 1.5 & 0.02 &  0.007 &  0.0006 \\
 $\sigma _{gT,1}\left(\tilde{f}_1\right)\;({\rm fb})$ & $3.55$ &  $1.38$ &  $1.44$ &  $1.34$ \\
 $\hat{\sigma} _{gT,1}\left(\tilde{f}_1\right)\;({\rm fb})$ & $3.52$ &  $1.37$ &  $1.43$ &  $1.33$ \\
\hline
 $f_{2,3}^U\;({\rm TeV}^{-4})$ & 14.0 & 0.114 & 0.030 &  0.004 \\
 $\tilde{f}_{2,3}\;({\rm TeV}^{-4})$ & 3 & 0.03 &  0.012 &  0.0012 \\
 $\sigma _{gT,2}\left(\tilde{f}_2\right)\;({\rm fb})$ & $3.83$ &  $0.956$ &  $1.34$ &  $1.78$ \\
 $\hat{\sigma} _{gT,2}\left(\tilde{f}_2\right)\;({\rm fb})$ & $3.82$ &  $0.955$ &  $1.34$ &  $1.78$ \\
 $\sigma _{gT,3}\left(\tilde{f}_3\right)\;({\rm fb})$ & $4.12$ & $0.924$ &  $1.27$ &  $1.61$ \\
 $\hat{\sigma} _{gT,3}\left(\tilde{f}_3\right)\;({\rm fb})$ & $4.10$ & $0.921$ &  $1.26$ &  $1.61$ \\
\hline
\end{tabular}
\caption{The cross-sections of the SM contribution and the NP contributions~\cite{Yang:2023gos}.
The NP contributions are cross-sections when the operator coefficients are $\tilde{f}_i$, $\tilde{f}_i$ used in the simulation are also shown.
The cross-sections after $N_j$ cut are denoted as $\hat{\sigma}$, which are also shown.
The $f_i^U$ are partial wave unitarity bounds on the operator coefficients.}
\label{table:coefficientscan}
\end{table}

To compare with the traditional event selection strategy, in this paper, we use the same events generated as in Ref.~\cite{Yang:2023gos}.
The events are generated using MC simulation with the \verb"MadGraph5_aMC@NLO" toolkit~\cite{Alwall:2014hca,Christensen:2008py,Degrande:2011ua}, where the standard cuts are set as the default.
The parton shower is applied using \verb"Pythia8"~\cite{Sjostrand:2014zea} with default settings.
A fast detector simulation is performed using \verb"Delphes"~\cite{deFavereau:2013fsa} with the muon collider card.
The data cleaning and preparation phase is applied using \verb"MLAnalysis"~\cite{Guo:2023nfu}.
The ANN is constructed and trained using the \verb"Keras" with the \verb"TensorFlow" backend~\cite{Abadi:2016kic}.
The events for the NP are generated with one operator at a time.
The operator coefficients are set as the same as Ref.~\cite{Yang:2023gos}.
As an EFT, the SMEFT is only valid under a certain energy scale.
One of the signals that the SMEFT is no longer valid is the violation of the unitarity~\cite{Lee:1960qv,Froissart:1961ux,Passarino:1990hk}, and the partial wave unitarity is often used in the phenomenological studies of the SMEFT to check whether the SMEFT is valid, which can sets bounds on the operator coefficients~\cite{Corbett:2014ora,Layssac:1993vfp,Corbett:2017qgl,Gomez-Ambrosio:2018nxl,Perez:2018kav,Almeida:2020ylr,Kilian:2018bhs}.
The operator coefficients used in the MC simulation are within the constraints set by the partial wave unitarity bounds~\cite{Yang:2023gos}.
The partial wave unitarity bounds~(denoted as $f_i^U$) are listed in Table~\ref{table:coefficientscan}.
Denoting $\sigma _{\rm SM}$ as the cross-section of the SM contribution, and $\sigma_{g T,i}\left(\tilde{f}_i\right)$ as NP contributions with the operator coefficients to be $\tilde{f}_i$, respectively, the cross-sections and operator coefficients $\tilde{f}_i$ are listed in Table~\ref{table:coefficientscan}.

In \verb"Delphes" we use the fast jet finder with anti-$k_T$ algorithm, and with $R=0.5$, $p_T^j > 20\;{\rm GeV}$, where $R$ is cone radius, and $p_T^j$ are the transverse momenta of jets.
After the events are generated, we require that the final states to have at least two jets.
This requirement is denoted as the $N_j$ cut, the cross-sections after $N_j$ cut are denoted as $\hat{\sigma}_{\rm SM}$ and $\hat{\sigma}_{g T,i}\left(\tilde{f}_i\right)$ which are also listed in Table~\ref{table:coefficientscan}.
Then an eight dimensional vector is made to represent each event such that $\vec{v}=(p^{(1)}_t,p^{(1)}_x,p^{(1)}_y,p^{(1)}_z,p^{(2)}_t,p^{(2)}_x,p^{(2)}_y,p^{(2)}_z)$, where $p^{(1)}$ and $p^{(2)}$ are the four momenta of hardest and second hardest jets. 
Note that $p_t^{(1)}$ and $p_t^{(2)}$ are the energies of jets in this paper.
In the following, we consider only the 8-dimensional vectors described above, ignoring the physical meaning behind them. 

\begin{table}[htbp]
\centering
\begin{tabular}{c|c|c|c|c}
\hline
 $\sqrt{s}$ & $\bar{v}_1$ & $\bar{v}_2$ & $\bar{v}_3$ & $\bar{v}_4$ \\
 TeV & GeV & GeV & GeV & GeV \\
\hline
$3$ & $271.00$ & $-0.14$ & $-0.19$ & $0.48$ \\
$10$ & $576.24$ & $-0.01$ & $0.14$ & $3.00$ \\
$14$ & $709.14$ & $-0.21$ & $-0.01$ & $0.57$ \\
$30$ & $1053.94$ & $0.15$ & $0.01$ & $4.64$ \\
 \hline
 $\sqrt{s}$ & $\bar{v}_5$ & $\bar{v}_6$ & $\bar{v}_7$ & $\bar{v}_8$ \\
 TeV & GeV & GeV & GeV & GeV \\
\hline
 $3$ & $110.13$ & $0.06$ & $0.04$ & $0.33$ \\
 $10$ & $225.00$ & $-0.04$ & $0.08$ & $1.45$ \\
 $14$ & $275.42$ & $-0.02$ & $-0.03$ & $0.03$ \\
 $30$ & $398.96$ & $-0.10$ & $-0.13$ & $1.74$ \\
\hline
$\sqrt{s}$ & $\epsilon_1$ & $\epsilon_2$ & $\epsilon_3$ & $\epsilon_4$ \\
 TeV & GeV & GeV & GeV & GeV \\
\hline
$3$ & 226.27 & 94.94 & 94.65 & 326.25 \\
$10$ & 665.88& 109.03 & 108.73 & 886.88 \\
$14$ & 879.40& 111.35 & 112.47 & 1118.43 \\
$30$ & 1504.96 & 120.43 & 122.45 & 1829.17 \\
\hline
$\sqrt{s}$ & $\epsilon_5$ & $\epsilon_6$ & $\epsilon_7$ & $\epsilon_8$ \\
 TeV & GeV & GeV & GeV & GeV \\
\hline
$3$ & 99.21 & 42.48 & 42.84 & 135.12 \\
$10$ & 284.41 & 48.77 & 48.20& 355.99 \\
$14$ & 371.30& 50.10& 52.11 & 456.52 \\
$30$ & 603.26 & 56.48 & 56.66 & 718.74 \\
\hline
\end{tabular}
\caption{The means $\bar{v}_i$ and standard deviations $\epsilon _i$ of the components of the vectors $\vec{v}$ over the training data sets.}
\label{table:Z-Scoredata}
\end{table}

For the SM contribution, we generate $1000000$ events for each c.m. energy, taking the case of $\sqrt{s}=3\;{\rm TeV}$ as an example, $832056$ events are left after the $N_j$ cut, of which $400000$ events consist the training data set, $100000$ events consist the validation data set, and rest events consist the test data set.
For each operator, $300000$ events are generated, and all the events after the $N_j$ cut are used as the test data set.
The numbers of events after the $N_j$ cut for the NP are generally more than $297000$.
Before the data sets are fed with the AE, a z-score standardization~\cite{Donoho_2004} is applied, i.e. the vectors $\vec{v}^n$ are replaced by $\vec{x}^{in,n}$ such that $x_i^{in,n} = \left(v_i^n-\bar{v}_i\right)/\epsilon_i$, where $x_i^{in,n},v_i^n$ are the $i$-th components of the $\vec{x}^{in,n}$ and $\vec{v}^n$ vectors, $\bar{v}_i$ and $\epsilon_i$ are the mean value and standard deviation of the $i$-th component over the training data set.
$\bar{v}_i$ and $\epsilon_i$ used in this paper are listed in Table~\ref{table:Z-Scoredata}.

The AE is to reproduce the input vectors, therefore the labels of the training data sets are just as same as the input of the data sets.

\subsection{\label{sec:42}Structure of the AE network}

Since the events are represented by eight dimensional vectors, the number of neurons in the input layer of the encoder, and in the output layer of the decoder are both eight.
The number of layers and the number of neurons in each layer is depicted in Fig.~\ref{fig:aestructure}.
In this paper, we use a densely connected ANN, taking the hidden layer $x_i^1$ as an example, values at neural $x_i^1$ can be calculated as $\vec{x}^1 = g^{{\rm in},1}(W^{{\rm in},1}\vec{x}^{\rm in}+\vec{b}^{\rm in})$, where $W^{i,j}$ is the weight matrix stored in the links between $i$-th and $j$-th layers, and $\vec{b}^{i}$ is the basis vector stored in the neurons in $i$-th layer, and $g^{i,j}$ is the activation function between $i$-th and $j$-th layers.
In this paper, we use the ``LeakyReLU'' function~\cite{Graves2012SupervisedSL} between the layers,
\begin{equation}
\begin{split}
&g(x)=\left\{
\begin{array}{cc}
x,& x>0,\\
\alpha x,& x\le 0,\\
\end{array}
\right.
\end{split}
\label{eq.activate}
\end{equation}
where $\alpha=0.01$.
Specifically, $g^{in,1},g^{1,2}$, $g^{latent,3}$, and $g^{3,4}$ are LeakyReLU functions, $g^{2,latent}$ and $g^{4,out}$ are linear activation functions.
The loss function defines how well the input vectors can be reproduced by the AE.
In this paper, we use the mse as the loss function.

Since it is expected that, the reason AE to be able to distinguish between the SM and NP is because the AE is able to find the patterns of the SM while being unable to learn the patterns of NP. 
Therefore we need the AE to have weak generalization properties and only reproduce events of the SM accurately, and smaller $k$ the better performance of AEAD is expected. 
To investigate the effect of the dimension of the latent space, four cases are considered, they are $k=1,2,3$ and $4$.

\subsection{\label{sec:43}Early stopping}

\begin{figure}[htbp]
\centering
\includegraphics[width=0.48\hsize]{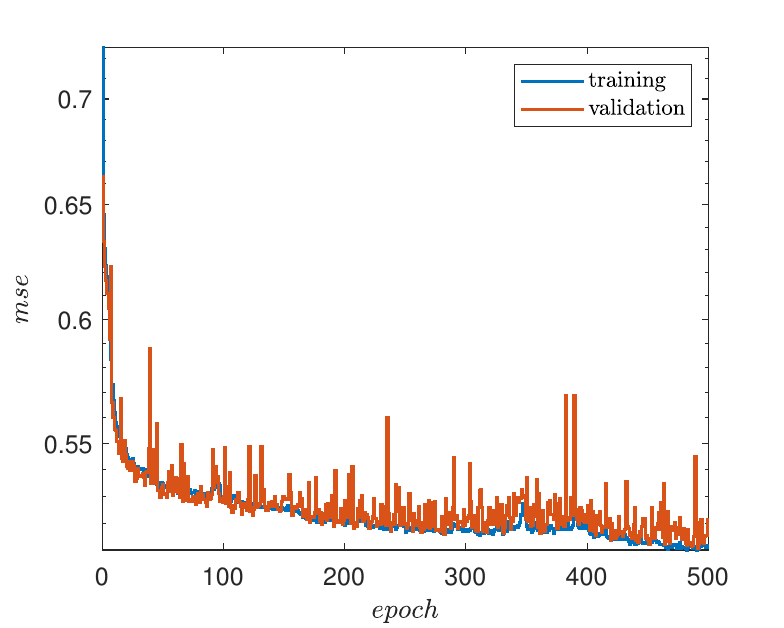}
\includegraphics[width=0.48\hsize]{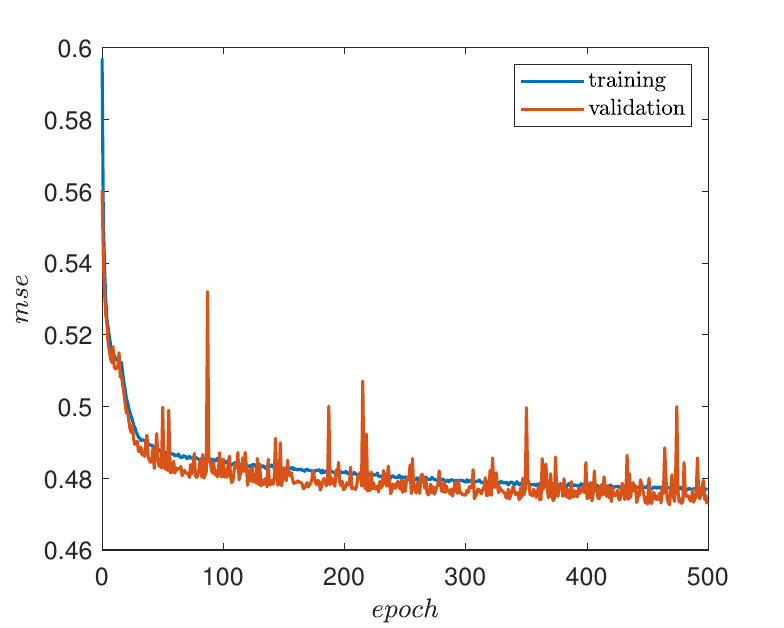}
\includegraphics[width=0.48\hsize]{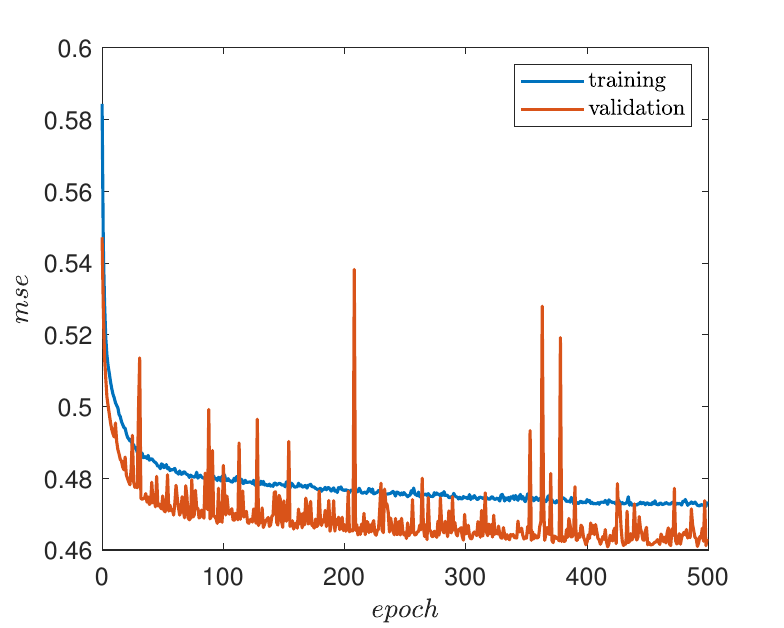}
\includegraphics[width=0.48\hsize]{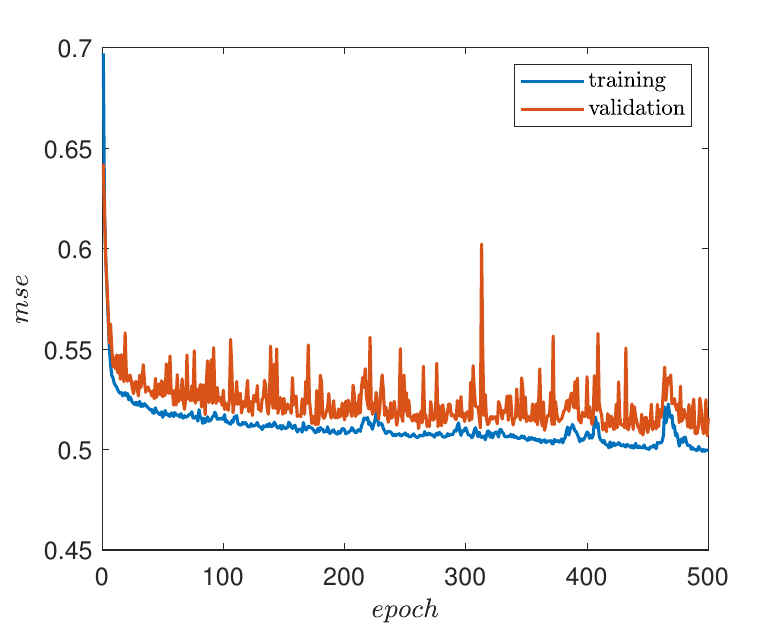}
\caption{\label{fig:learningcurve}
The learning curves in the training phase for $k=1$. 
The top-left panel corresponds to $\sqrt{s} = 3\;\rm TeV$, the top-right panel corresponds to $\sqrt{s} =10\;\rm TeV $, the bottom-left panel corresponds to $\sqrt{s} =14\;\rm TeV $, and the bottom-right panel corresponds to $\sqrt{s} =30\;\rm TeV $.}
\end{figure}

\begin{figure}[htbp]
\centering
\includegraphics[width=0.48\hsize]{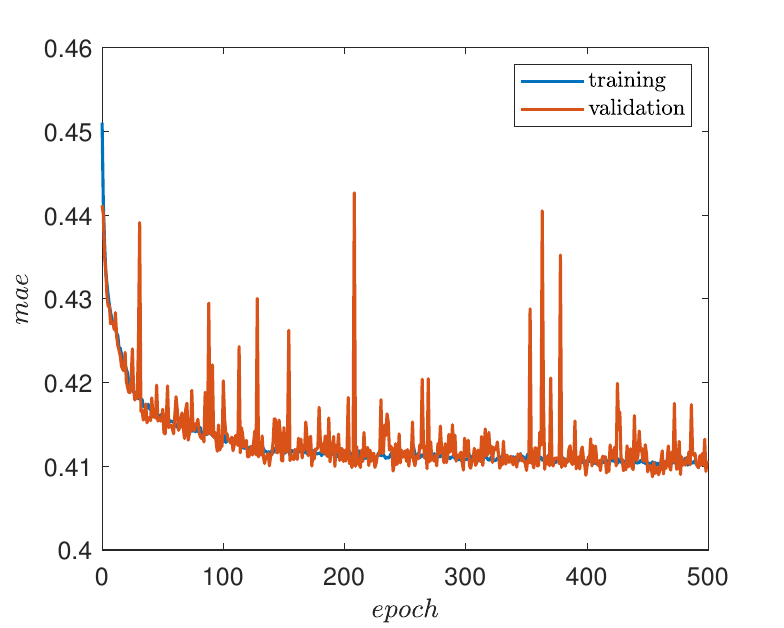}
\caption{\label{fig:maecurve}
The maes of training and validation data sets at $\sqrt{s} =14\;\rm TeV $.}
\end{figure}

Overfitting is a situation where the model performs well on the training set but relatively poorly on the test set.
In that case, the model is weak in predicting unknown data. 
One of the methods to avoid overfitting is early stop.

The process of training the model is the process of updating the model parameters~(i.e. $W^{i,j}$ and $\vec{b}^i$) through learning.
In the training phase, the data set is divided into a training data set and a validation data set, and only the training data set is used to update the parameters.
During the training process, the errors for the training set and validation set gradually decreases, and after reaching a critical point, the errors for the training set continue to decrease and the ones for the validation set start to increase.
Early stopping is to prevent overfitting by stopping the training before the critical point, i.e., the number of iterations is truncated.

We use early stopping method to avoid overfitting in this paper.
As an example, the mses for the training data set and the validation data set as functions of number of epochs for $k=1$ are shown in Fig.~\ref{fig:learningcurve}.
We stop when the mean of mse of last $50$ epochs of the validation set is larger than then the one of the last $100$ epochs, which is checked every $100$ epochs.
In the training, we preserve the networks by training to $500$ epochs while keeping only the one with the smallest mses for validation data sets. 
So when mse of the validation data set plateaus early, or starts to rise, the result with the smallest mse will not be missed.

It has been noticed that the mse of the validation data set is smaller than the one of the training data set, for example the case at $\sqrt{s}=14\;{\rm TeV}$ in Fig.~\ref{fig:learningcurve}.
It is found that whether or not validation is better than training occurs mainly depends on the segmentation of the training and validation data sets. 
Meanwhile, if one looks at the mean absolute error~(mae) of the training and validation sets, which is defined as $\sum _n^N\sum _i^8\left|x^{in,n}_i-X^{out,n}_i\right|/N$, one will find that the two are always close to each other, for example as the case at $\sqrt{s}=14\;{\rm TeV}$ shown in Fig.~\ref{fig:maecurve}. 
Since mse is more sensitive to anomalous samples, it can inferred that, the above phenomenon is due to the fact that there are events in the SM data set that are more anomalous relative to the rest of the SM events, and that where they are partitioned determines the mse. 
Normally, this is a situation more appropriate to use mae as a loss function, however, since mse is more sensitive to anomalous samples and our goal is not to reproduce the SM events but to find anomalies, we still use mse as a loss function.

\subsection{\label{sec:44}Distribution of the anomaly score}

\begin{figure}[htbp]
\centering
\includegraphics[width=0.48\hsize]{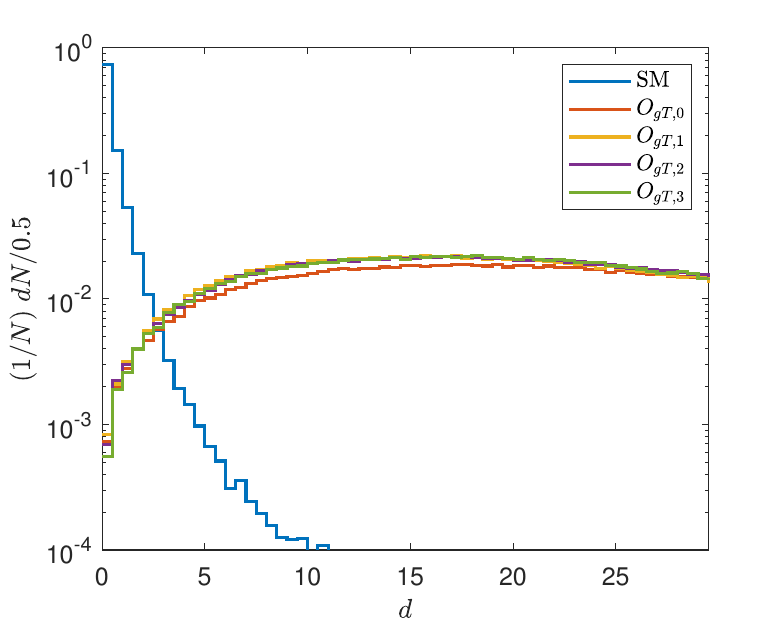}
\includegraphics[width=0.48\hsize]{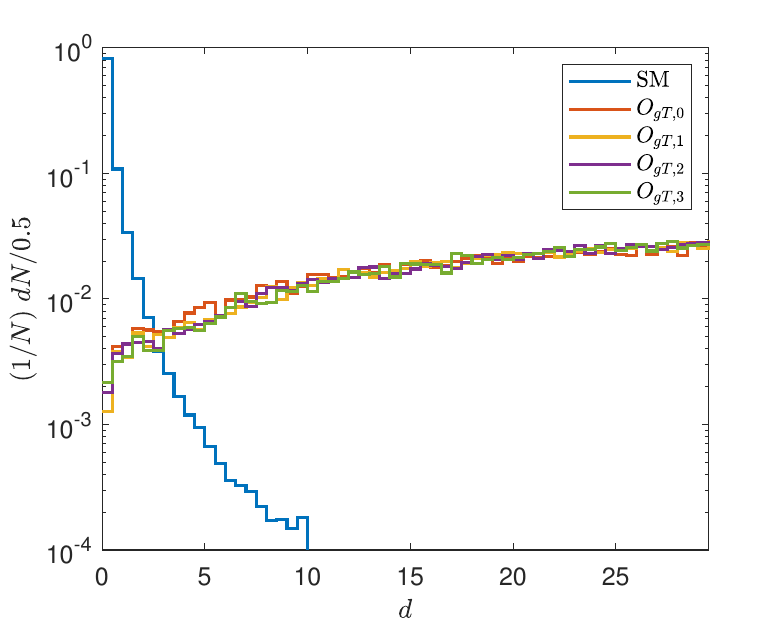}
\includegraphics[width=0.48\hsize]{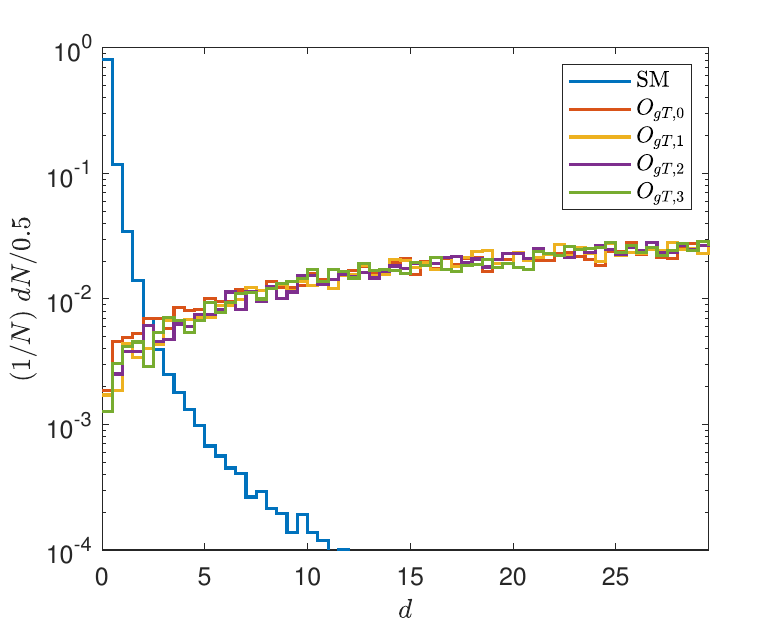}
\includegraphics[width=0.48\hsize]{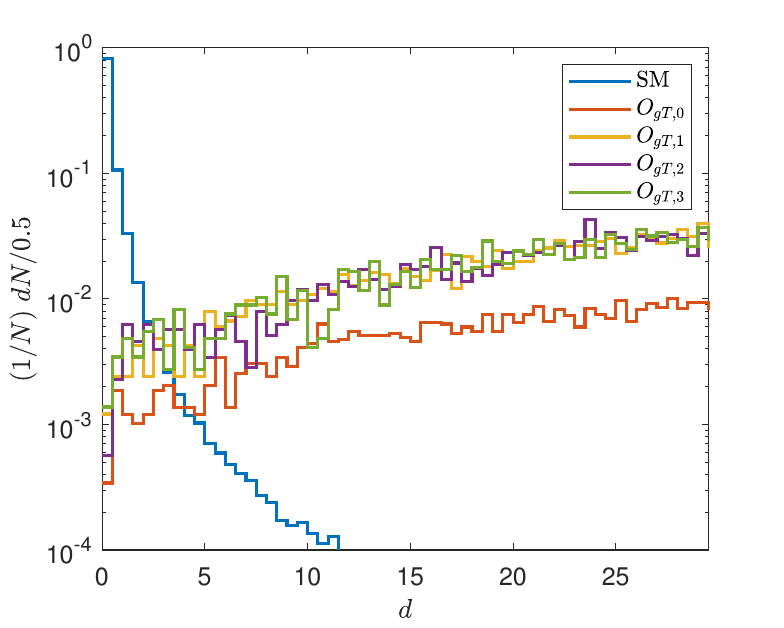}
\caption{\label{fig:msecompare}
The normalized distributions of $d$ when $k=1$ for the SM events and for the NP events in the test data sets.
The top-left panel corresponds to $\sqrt{s} = 3\;\rm TeV$, the top-right panel corresponds to $\sqrt{s} =10\;\rm TeV $, the bottom-left panel corresponds to $\sqrt{s} =14\;\rm TeV $, and the bottom-right panel corresponds to $\sqrt{s} =30\;\rm TeV $.}
\end{figure}

As described in the previous section, the AEAD uses how well the AE can reproduce the input as the anomaly score. 
That is, one can use $d$ as the anomaly score.
Note that $d$ is defined using the data after z-score standardization, therefore is dimensionless.

In this subsection and in the following, we concentrate on the test data sets.
As an example, the normalized distributions of $d$ when $k=1$ are shown in Fig.~\ref{fig:msecompare}.
It can be seen that the $d$ for the SM background at different energies are generally small, thanks to the well trained AE.
Meanwhile, the $d$ for the NP signals are generally larger, and the larger the $\sqrt{s}$, the larger $d$.
From the distributions it can be conclude that $d$ can provide a good discriminate ability to select the NP signals as expected.

\subsection{\label{sec:45}Latent space distribution}

\begin{figure}[htbp]
\centering
\includegraphics[width=0.48\hsize]{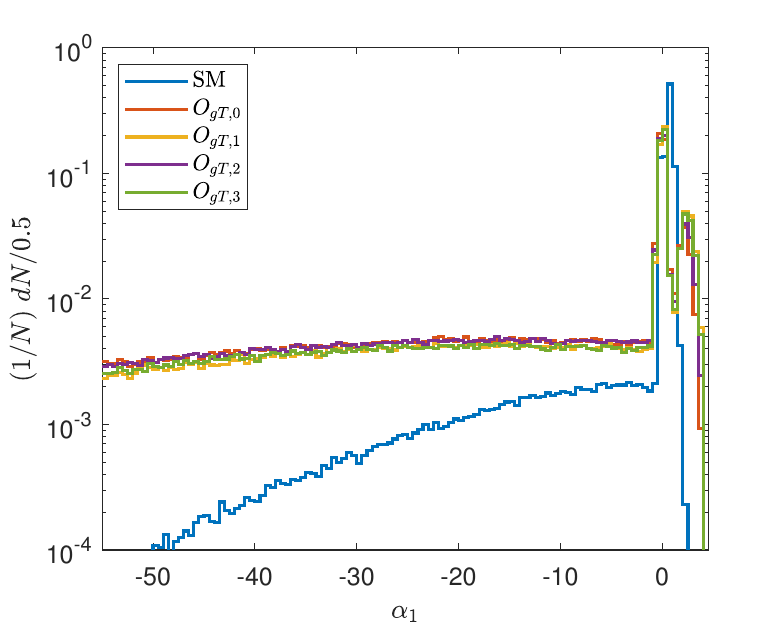}
\includegraphics[width=0.48\hsize]{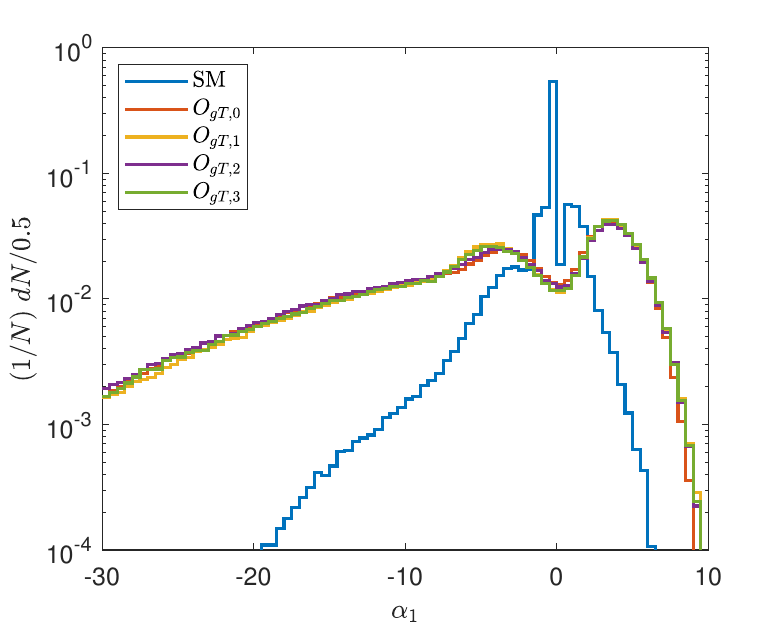}
\includegraphics[width=0.48\hsize]{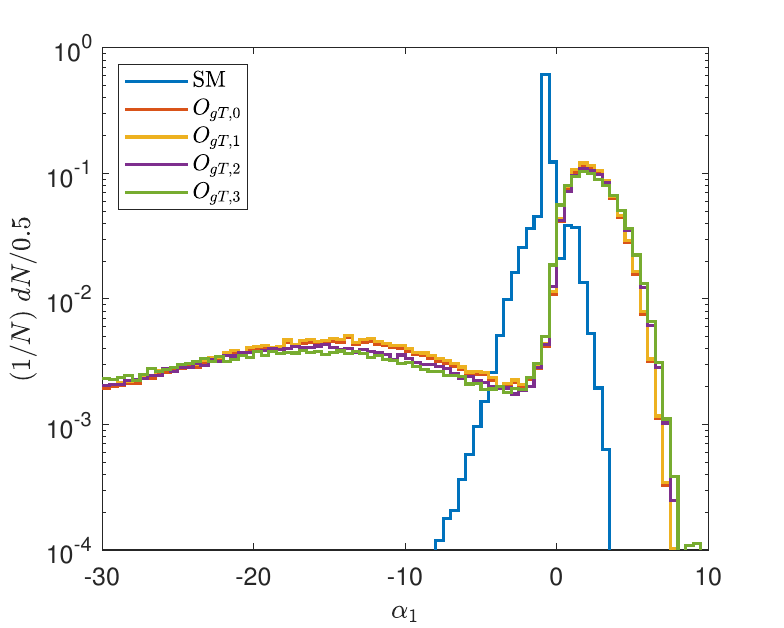}
\includegraphics[width=0.48\hsize]{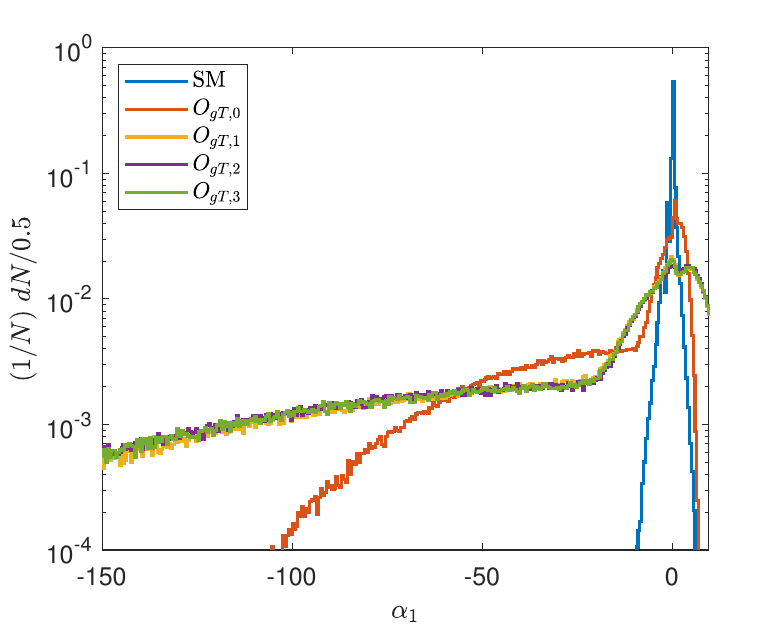}
\caption{\label{fig:Laterntspacek1}
The distributions of the test data set events in the latent space when $k=1$.
The top-left panel corresponds to $\sqrt{s} = 3\;\rm TeV$, the top-right panel corresponds to $\sqrt{s} =10\;\rm TeV $, the bottom-left panel corresponds to $\sqrt{s} =14\;\rm TeV $, and the bottom-right panel corresponds to $\sqrt{s} =30\;\rm TeV $.} 
\end{figure}

\begin{figure*}[htbp]
\centering
\includegraphics[width=0.24\hsize]{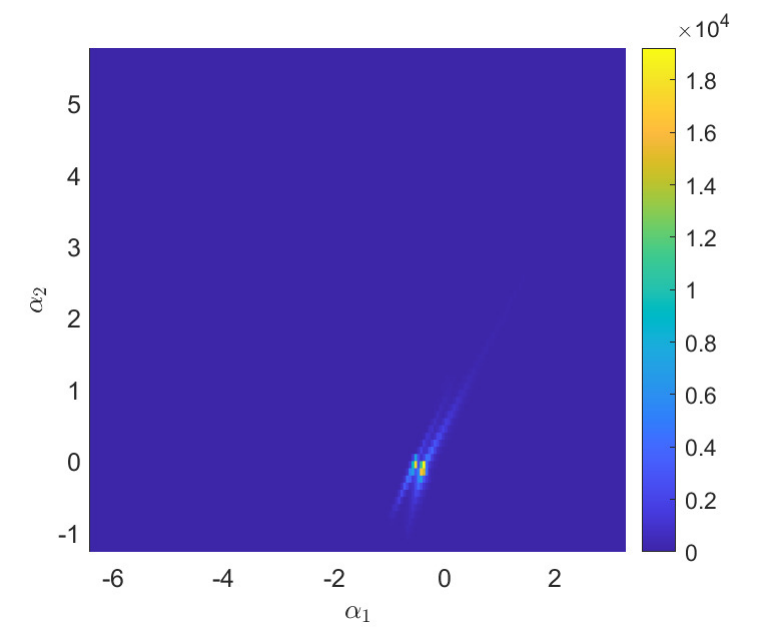}
\includegraphics[width=0.24\hsize]{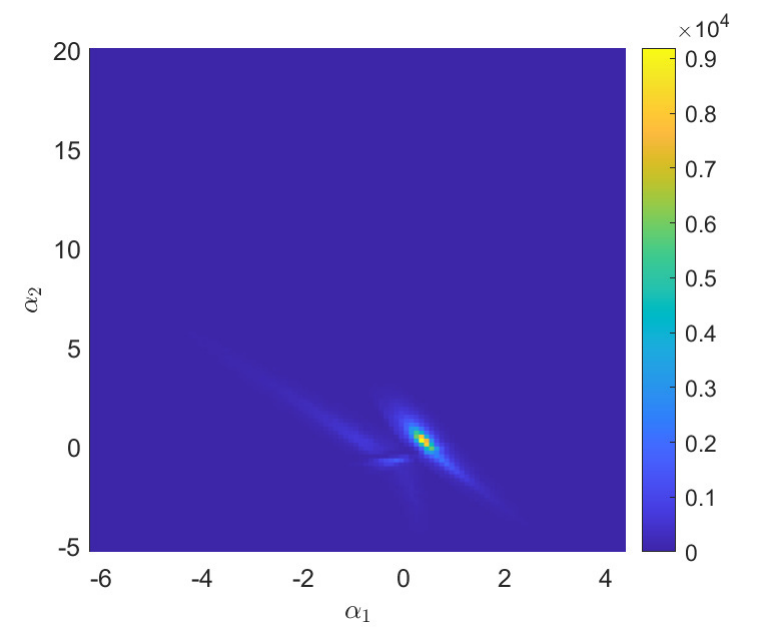}
\includegraphics[width=0.24\hsize]{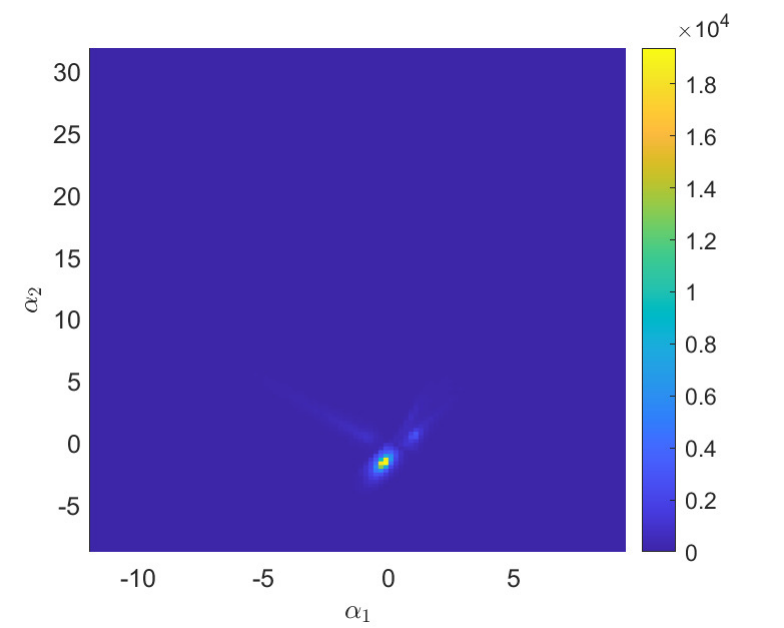}
\includegraphics[width=0.24\hsize]{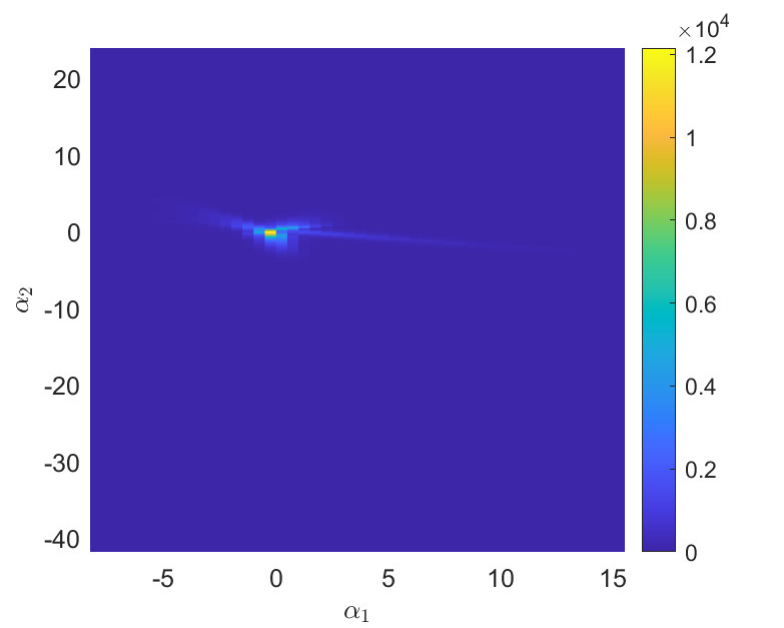}
\includegraphics[width=0.24\hsize]{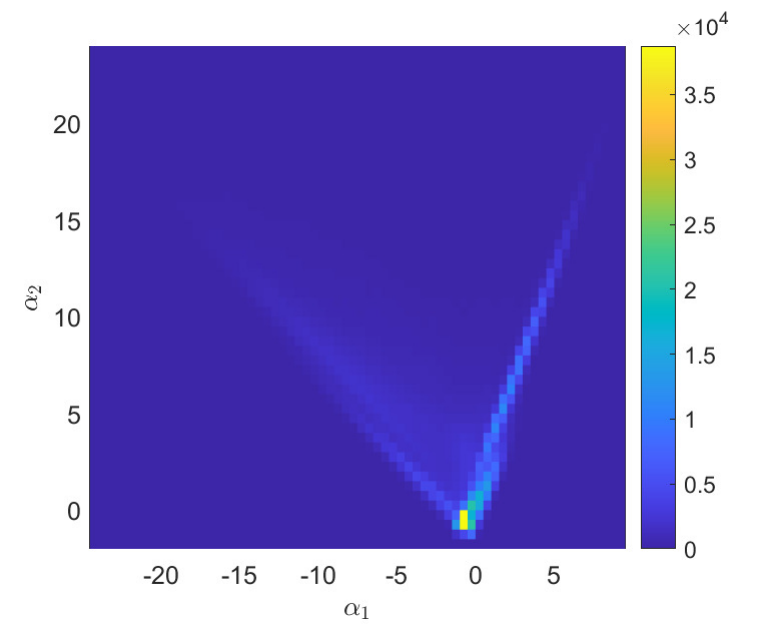}
\includegraphics[width=0.24\hsize]{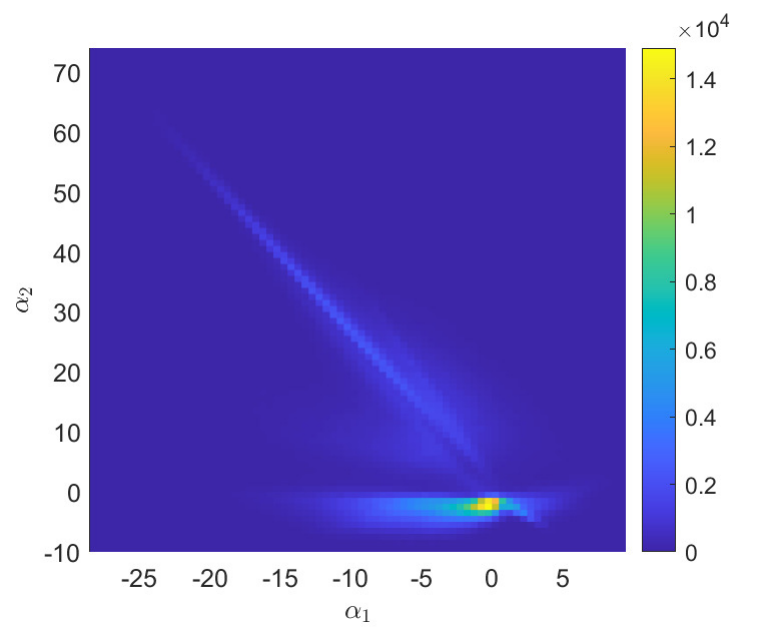}
\includegraphics[width=0.24\hsize]{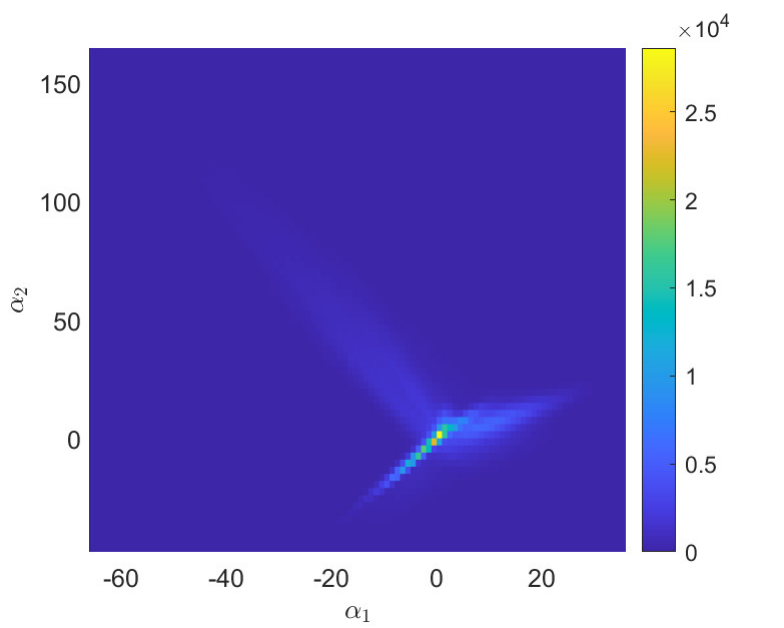}
\includegraphics[width=0.24\hsize]{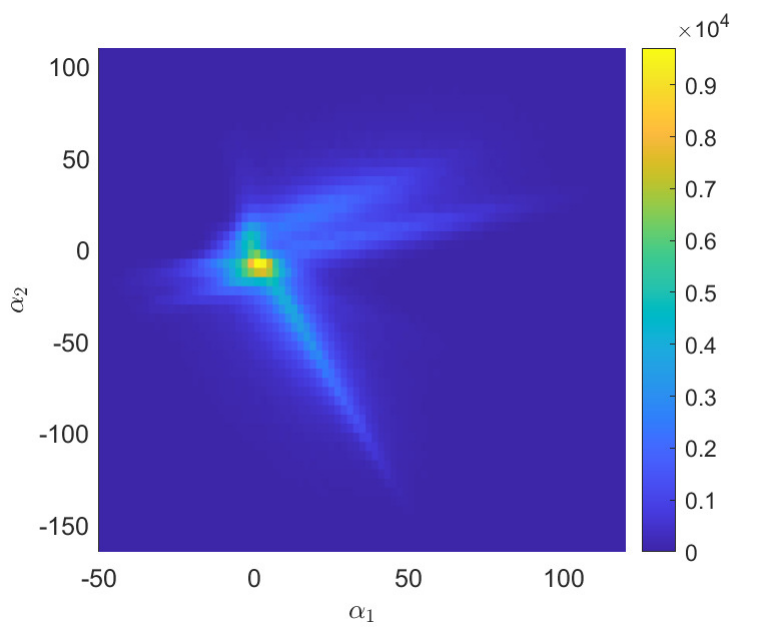}
\caption{\label{fig:Laterntspacek2}
Same as Fig.~\ref{fig:Laterntspacek1} but for $k=2$.
The panels in the first row are for the SM test data set, and those in the second row are for the $O_{gT,0}$ test data sets.
The first, second, third and the fourth columns correspond to $\sqrt{s}=3\;{\rm TeV}$, $10\;{\rm TeV}$, $14\;{\rm TeV}$ and $30\;{\rm TeV}$, respectively.} 
\end{figure*}

It is known that AE can also be used as a data dimensionality reduction algorithm, a scheme that uses AE for classification uses AE as a data preparation stage.
In combination with AE, other classification algorithms or AD algorithms can be applied in the latent space, i.e., the space consisting of the $\alpha _i$ values that are in the middle layer. 
Since the AE is trained to approximately reproduce events of the SM, this means that $\alpha_i$ contain the major information needed to reconstruct the events. 
For this to happen, there are hidden relationships between the components of the vectors input to the AE, as a result, the components of the input vectors are not independent of each other.
After dimension reduction, the events can be represented by a smaller number~($k$ in our case) of variables.
Therefore, with the help of latent space, the features of the SM and NP are more easily presented visually.

The distributions of the events in the latent spaces when $k=1$ are shown in Fig.~\ref{fig:Laterntspacek1}, and the case for $k=2$ are shown in Fig.~\ref{fig:Laterntspacek2}.
It can be seen that, in the latent spaces, the NP events already distribute differently from the SM events.
Due to the nonlinearity of the activation functions~(segmented functions are used as the activation functions), different regions in the latent space often represent different functions that have been imposed to $\alpha _i$ in the decoder, and therefore represent different relationships among the components of the input/output vectors.
The fact that the SM and NP events distributed differently in the latent space reflects the conjecture that, since the hidden relationships between the components are obtained by having the AE trained on the SM events, the events of the NP are not described by these relationships. 
This explains why $d$ can be used to search for the NP signals.

\subsection{\label{sec:46}Cut efficiency}

Since there is no interference between the SM and NP, the cross-section after cut can be written as,
\begin{equation}
\begin{split}
&\sigma(f_i) = \varepsilon _{\rm SM}\hat{\sigma} _{\rm SM} + \varepsilon _{O_{gT,i}}\frac{f_i^2}{\tilde{f}_i^2} \hat{\sigma}_{\mathrm{g} T,i}\left(\tilde{f}_i\right),
\end{split}
\label{eq.crosssection}
\end{equation}
where $\hat{\sigma} _{\rm SM}$ and $\hat{\sigma}_{\mathrm{g} T,i}\left(\tilde{f}_i\right)$ are listed in Table~\ref{table:coefficientscan}, $\varepsilon _{\rm SM}$ is the cut efficiency of selecting events with $d>d_{th}$ for the SM, and $\varepsilon _{O_{gT,i}}$ are cut efficiencies for the NP signals.
Note that, $\varepsilon$ does not include the effect of the $N_j$ cut~(denoted as $\varepsilon _{N_j}$), which is already included in $\hat{\sigma}=\varepsilon _{N_j}\sigma $.
$\varepsilon _{\rm SM}$ and $\varepsilon _{O_{gT,i}}$ are functions of $d_{th}$, and to facilitate the study of expected coefficient constraints, we fit the cut efficiencies using rational functions,
\begin{equation}
\begin{split}
&\varepsilon _{SM}(d_{th})=\frac{1+a_1 d_{th}+a_5d_{th}^2}{a_2+a_3d_{th}+a_4d_{th}^2+a_6d_{th}^3},\\
&\varepsilon(d_{th})=\frac{1+a_1 d_{th}}{a_2+a_3d_{th}+a_4d_{th}^2}.
\end{split}
\label{eq.cutefficiencyansatz}
\end{equation}
For the case of $\varepsilon_{SM}$ when $k=1$ and $\sqrt{s}=3\;{\rm TeV}$, we use a rational function with six parameters, and for the other cases we use a rational function with four parameters.

\begin{figure}[htbp]
\centering
\includegraphics[width=0.48\hsize]{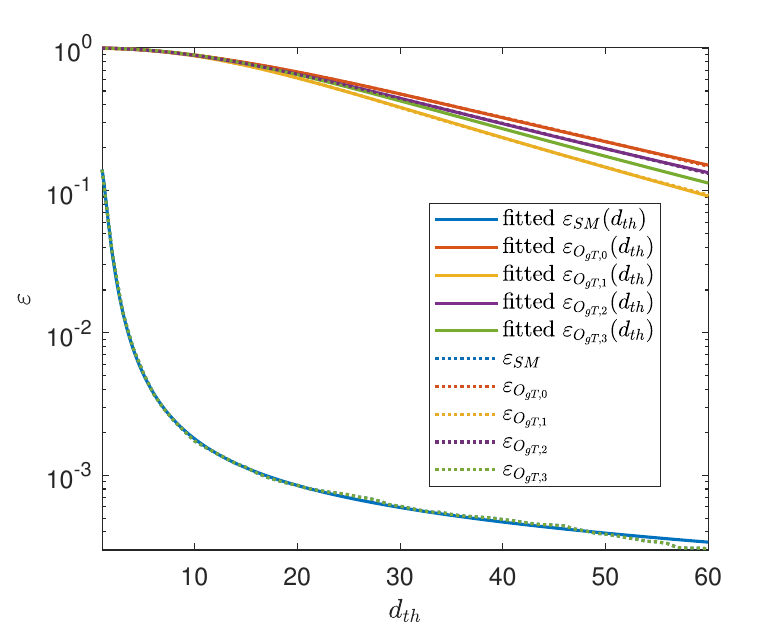}
\includegraphics[width=0.48\hsize]{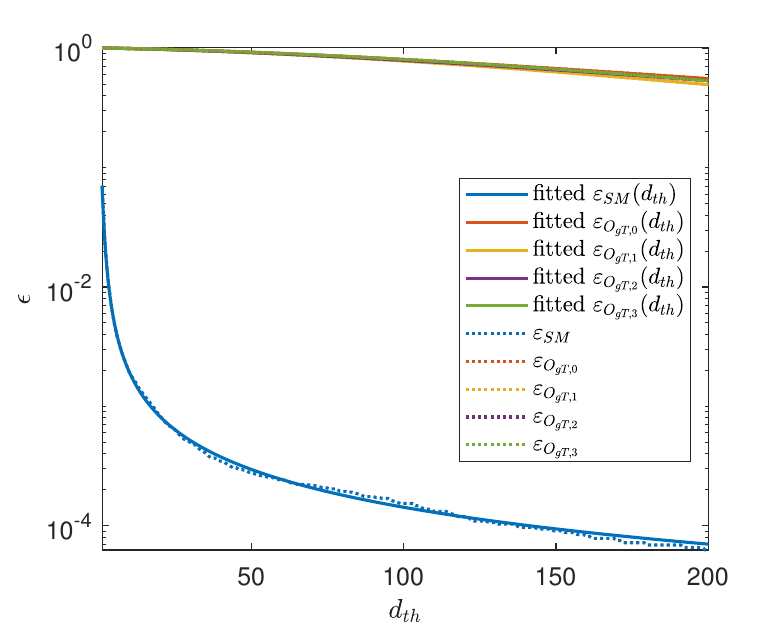}\\
\includegraphics[width=0.48\hsize]{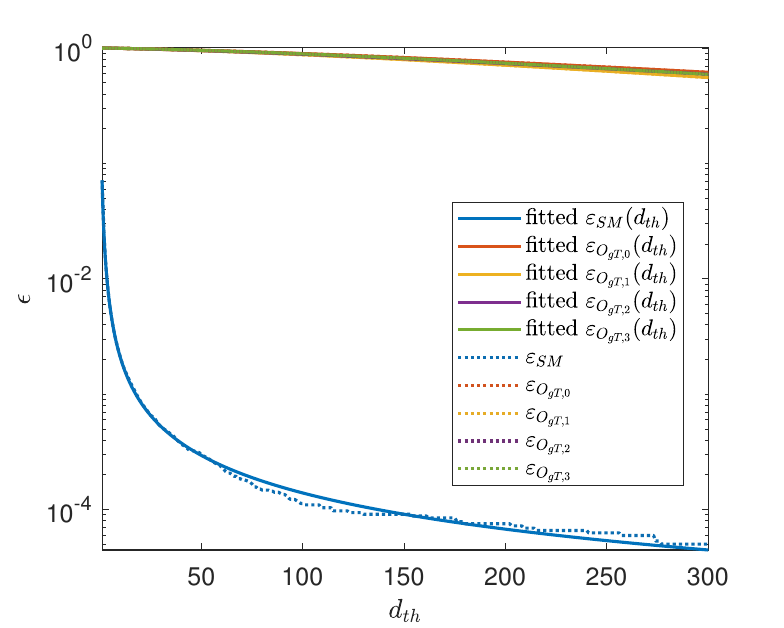}
\includegraphics[width=0.48\hsize]{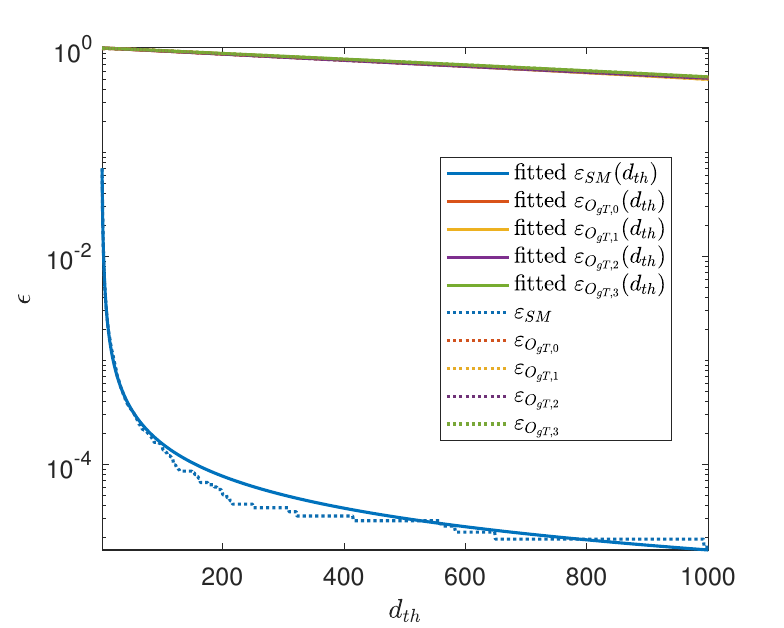}
\caption{\label{fig:cutefficiency}
$\varepsilon $ as functions of $d_{th}$ compared with the fitted $\varepsilon (d_{th})$. 
The top-left panel corresponds to $\sqrt{s} = 3\;\rm TeV$, the top-right panel corresponds to $10\;\rm TeV $, the bottom-left panel corresponds to $14\;\rm TeV $, and the bottom-right panel corresponds to $30\;\rm TeV $.}
\end{figure}

As an example, the results of the fits in the case of $k=1$ are shown in Fig.~\ref{fig:cutefficiency}.
It can be seen that, $\varepsilon_{SM}$ is much smaller than $\varepsilon _{O_{gT,i}}$, indicates that the event selection strategy using $d_{th}$ can be used to suppress the background.

\subsection{\label{sec:47}Expected constraints on the operator coefficients}


Usually, when the NP signals are not found, the task is to set constraints on the operator coefficients.
This can be done by using the statistical signal significance which is defined as~\cite{Cowan:2010js,ParticleDataGroup:2020ssz},
\begin{equation}
\begin{split}
&\mathcal{S}_{\rm stat}=\sqrt{2\left[\left(N_{\rm bg}+N_{s}\right) \ln \left(1+N_s / N_{\rm bg}\right)-N_s\right]},
\end{split}
\label{eq.stat}
\end{equation}
where $N_{\rm bg}$ is the number of background events, $N_s$ is the number of signal events.
Since there is no interference between the SM and NP, the number of events after cuts can be obtained by $N_s= \varepsilon _{O_{gT,i}} L\times \hat{\sigma} _{gT,i}$ and $N_{\rm bg}= \varepsilon _{\rm SM} L\times \hat{\sigma} _{\rm SM}$, where $\varepsilon$ is the cut efficiency, $\hat{\sigma}$ is the cross-section after $N_j$ cut, and $L$ is the luminosity.

\begin{figure}[htbp]
\centering
\includegraphics[width=0.48\hsize]{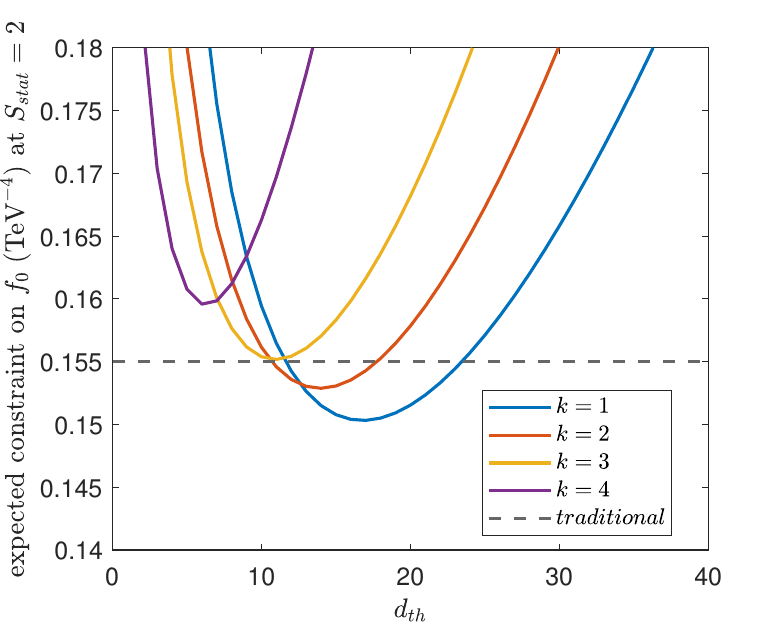}
\includegraphics[width=0.48\hsize]{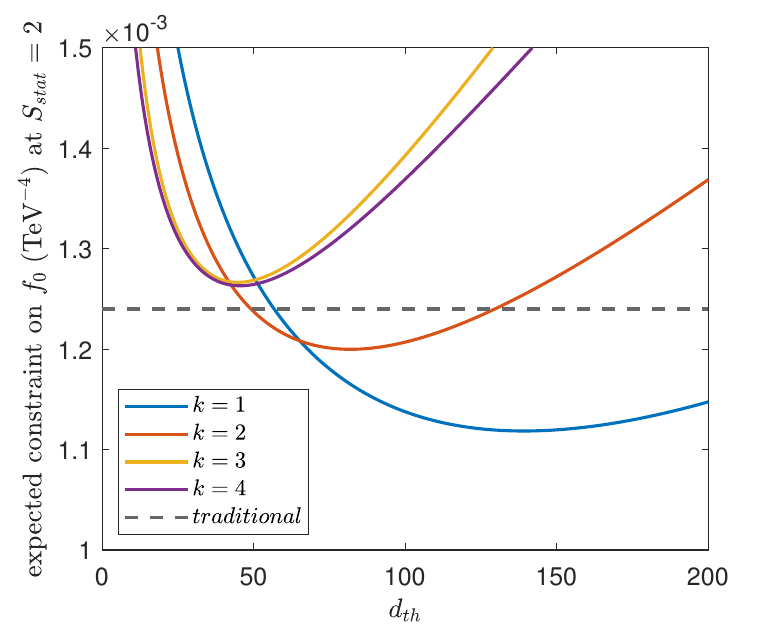}
\includegraphics[width=0.48\hsize]{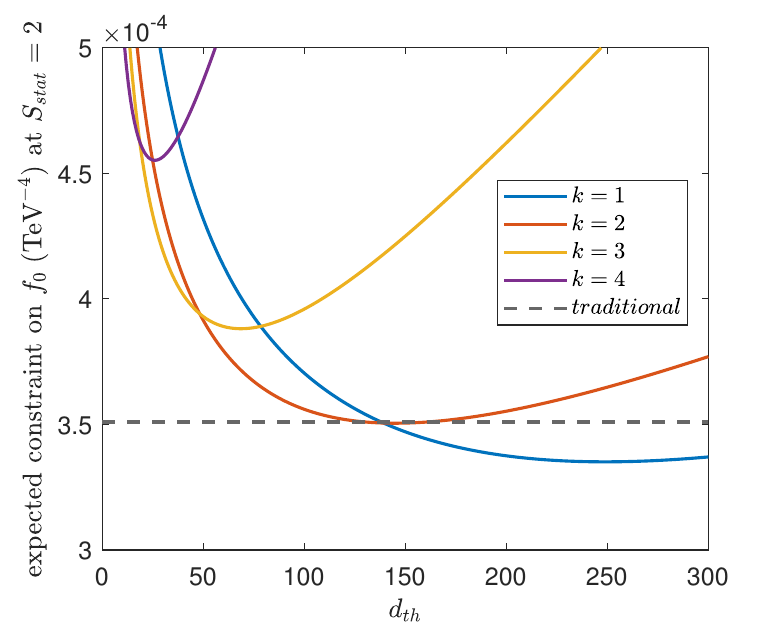}
\includegraphics[width=0.48\hsize]{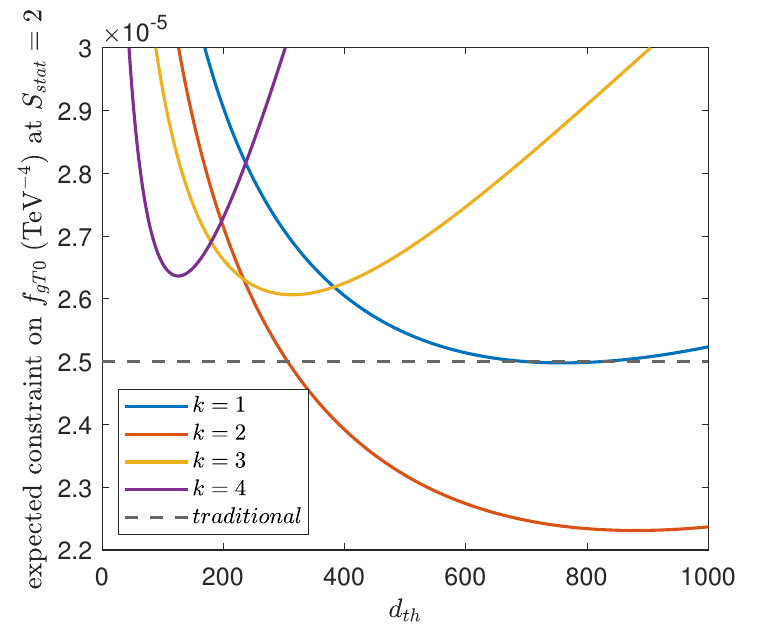}
\caption{\label{fig:constraints}
The expected constraints on $f_0$ at different energies as functions of $d_{th}$ when $\mathcal{S}_{stat}=2$ in the ``conservative'' case.
The top-left panel corresponds to $\sqrt{s} = 3\;\rm TeV$, the top-right panel corresponds to $\sqrt{s} =10\;\rm TeV $, the bottom-left panel corresponds to $\sqrt{s} =14\;\rm TeV $, and the bottom-right panel corresponds to $\sqrt{s} =30\;\rm TeV $.
}
\end{figure}

\begin{table}
\begin{center}
\begin{tabular}{c|c|c|c|c|c}
    \hline
    $ $ & \multirow{2}{*}{$k$}  & $3\;\mathrm{TeV}$ & $10\;\mathrm{TeV}$ & $14\;\mathrm{TeV}$ & $30\;\mathrm{TeV}$ \\
    $ $ &$ $ & $1\;{\rm ab}^{-1}$ & $10\;{\rm ab}^{-1}$ & $10\;{\rm ab}^{-1}$ & $10\;{\rm ab}^{-1}$ \\
    \hline
    \multirow{4}{*}{$d_{th}$} &$1$ & $16.8$ & $139.1$ & $248.8$ & $745.1$\\
    $ $ &$2$ & $13.9$ & $81.9$ & $144.9$ & $879.7$\\
    $ $ &$3$ & $10.9$ & $44.8$ & $68.7$ & $313.3$\\
    $ $ &$4$ & $6.3$ & $45.5$ & $26.1$ & $125.6$\\
    \hline
\end{tabular}
\end{center}
\caption{\label{tab:dthtochoose}
The values of $d_{th}$ which minimize the expected constraints on $f_{0}$ when $\mathcal{S}_{stat}=2$ in the ``conservative'' case.} 
\end{table}
The luminosities in the ``conservative'' case for the muon colliders at $\sqrt{s}=3\;{\rm TeV}$, $10\;{\rm TeV}$, $14\;{\rm TeV}$ and $30\;{\rm TeV}$ are $L=1\;{\rm ab}^{-1}$, $10\;{\rm ab}^{-1}$, $10\;{\rm ab}^{-1}$ and $10\;{\rm ab}^{-1}$, respectively~\cite{AlAli:2021let}.
Using the fitted $\varepsilon(d_{th})$, the expected constraints on the operator coefficients can be directly obtained.
The expected constraints for $\mathcal{S}_{\rm stat}=2$ in the ``conservative'' case are shown in Fig.~\ref{fig:constraints}.
It can be shown that, the cases for $k=1$ and $k=2$ can perform better than the traditional event selection strategy.
At $\sqrt{s}=3$, $10$ and $14$ TeV, $k=1$ works best, and at $\sqrt{s}=30$ TeV, $k=2$ works best.
In the following, we choose $d_{th}$ which minimize the expect constraints according to Fig.~\ref{fig:constraints}.
The results of $d_{th}$ are listed in Table~\ref{tab:dthtochoose}.

\begin{table}
\begin{center}
\begin{tabular}{c|c|c|c|c|c|c} 
    \hline
    $ $ & $ $ & \multicolumn{4}{c|}{$k=1$ $ $ $ $ $ $}& $k=2$ \\
    \hline
    $ $ & $\mathcal{S}_{stat}$ & $3\;\mathrm{TeV}$ & $10\;\mathrm{TeV}$ & $14\;\mathrm{TeV}$ & $30\;\mathrm{TeV}$ & $30\;\mathrm{TeV}$ \\ 
    $ $ & $ $ & $1\;{\rm ab}^{-1}$ & $10\;{\rm ab}^{-1}$ & $10\;{\rm ab}^{-1}$ & $10\;{\rm ab}^{-1}$ & $10\;{\rm ab}^{-1}$ \\ 
    \hline
    $ $       & $2$ & $<149$& $<1.11$& $<0.333$& $<0.0248$& $<0.0222$\\ 
    $|f_{0}|$ & $3$ & $<184$& $<1.36$& $<0.410$& $<0.0306$& $<0.0273$\\ 
    $ $       & $5$ & $<239$& $<1.77$& $<0.532$& $<0.0498$& $<0.0356$\\ 
    \hline
    $ $     & $2$ & $>0.804$& $>2.74$& $>3.70$& $>7.08$& $>7.28$\\ 
    $M_{0}$ & $3$ & $>0.764$& $>2.60$& $>3.51$& $>6.72$& $>6.92$\\ 
    $ $     & $5$ & $>0.715$& $>2.44$& $>3.29$& $>6.29$& $>6.47$\\ 
    \hline
    $ $       & $2$ & $<221$& $<1.75$& $<0.538$& $<0.0398$& $<0.0382$\\ 
    $|f_{1}|$ & $3$ & $<272$& $<2.14$& $<0.661$& $<0.0490$& $<0.0471$\\ 
    $ $       & $5$ & $<353$& $<2.78$& $<0.858$& $<0.0638$& $<0.0614$\\ 
    \hline
    $ $     & $2$ & $>0.729$& $>2.45$& $>3.28$& $>6.29$& $>6.36$\\ 
    $M_{1}$ & $3$ & $>0.692$& $>2.32$& $>3.12$& $>5.98$& $>6.04$\\ 
    $ $     & $5$ & $>0.649$& $>2.18$& $>2.92$& $>5.59$& $>5.65$\\ 
    \hline
    $ $       & $2$ & $<416$& $<3.09$& $<0.930$& $<0.0686$& $<0.0623$\\ 
    $|f_{2}|$ & $3$ & $<511$& $<3.79$& $<1.142$& $<0.0845$& $<0.0768$\\ 
    $ $       & $5$ & $<664$& $<4.92$& $<1.484$& $<0.1100$& $<0.1001$\\ 
    \hline
    $ $     & $2$ & $>0.622$& $>2.12$& $>2.86$& $>5.49$& $>5.63$\\ 
    $M_{2}$ & $3$ & $>0.591$& $>2.01$& $>2.72$& $>5.22$& $>5.34$\\ 
    $ $     & $5$ & $>0.554$& $>1.89$& $>2.55$& $>4.88$& $>5.00$\\ 
    \hline
    $ $       & $2$ & $<402$& $<3.12$& $<0.957$& $<0.0710$& $<0.0667$\\ 
    $|f_{3}|$ & $3$ & $<494$& $<3.83$& $<1.176$& $<0.0873$& $<0.0821$\\ 
    $ $       & $5$ & $<641$& $<4.97$& $<1.528$& $<0.1137$& $<0.1071$\\ 
    \hline
    $ $     & $2$ & $>0.628$& $>2.12$& $>2.84$& $>5.45$& $>5.53$ \\ 
    $M_{3}$ & $3$ & $>0.596$& $>2.01$& $>2.70$& $>5.17$& $>5.25$ \\ 
    $ $     & $5$ & $>0.559$& $>1.88$& $>2.53$& $>4.84$& $>4.91$ \\ 
    \hline
\end{tabular}
\end{center}
\caption{\label{tab:conservative}
Expected constraints on $f_i\;(10^{-3} \mathrm{TeV^{-4}})$ and $M_i\;(\mathrm{TeV})$ in the ``conservative'' case at muon colliders.} 
\end{table}
\begin{table}
\begin{center}
\begin{tabular}{c|c|c|c|c}
    \hline
    $ $ & $ $ & \multicolumn{2}{c|}{$k=1$ $ $}& $k=2$ \\
    \hline
    $ $ & $\mathcal{S}_{stat}$ & $14\;\mathrm{TeV}$ & $30\;\mathrm{TeV}$ & $30\;\mathrm{TeV}$ \\
    $ $ & $ $ & $20\;{\rm ab}^{-1}$ & $90\;{\rm ab}^{-1}$ & $90\;{\rm ab}^{-1}$ \\
    \hline
    $|f_{0}|$                     & $2$ & $<2.80$& $<0.142$& $<0.127$\\
    $(10^{-4} \mathrm{TeV^{-4}})$ & $3$ & $<3.43$& $<0.175$& $<0.156$\\
    $ $                           & $5$ & $<4.45$& $<0.226$& $<0.202$\\
    \hline
    $M_{0}$          & $2$ & $>3.87$& $>8.14$& $>8.37$\\
    $(\mathrm{TeV})$ & $3$ & $>3.67$& $>7.73$& $>7.95$\\
    $ $              & $5$ & $>3.44$& $>7.25$& $>7.46$\\
    \hline
    $|f_{1}|$                     & $2$ & $<4.51$& $<0.228$& $<0.219$\\
    $(10^{-4} \mathrm{TeV^{-4}})$ & $3$ & $<5.54$& $<0.280$& $<0.269$\\
    $ $                           & $5$ & $<7.18$& $<0.363$& $<0.349$\\
    \hline
    $M_{1}$          & $2$ & $>3.43$& $>7.23$& $>7.31$\\
    $(\mathrm{TeV})$ & $3$ & $>3.26$& $>6.87$& $>6.94$\\
    $ $              & $5$ & $>3.05$& $>6.44$& $>6.51$\\
    \hline
    $|f_{2}|$                     & $2$ & $<7.80$ & $<0.394$& $<0.357$\\
    $(10^{-4} \mathrm{TeV^{-4}})$ & $3$ & $<9.58$ & $<0.483$& $<0.438$\\
    $ $                           & $5$ & $<12.42$& $<0.626$& $<0.568$\\
    \hline
    $M_{2}$          & $2$ & $>2.99$& $>6.31$& $>6.47$\\
    $(\mathrm{TeV})$ & $3$ & $>2.84$& $>6.00$& $>6.14$\\
    $ $              & $5$ & $>2.66$& $>5.62$& $>5.76$\\
    \hline
    $|f_{3}|$                     & $2$ & $<8.04$ & $<0.407$& $<0.382$\\
    $(10^{-4} \mathrm{TeV^{-4}})$ & $3$ & $<9.86$ & $<0.499$& $<0.469$\\
    $ $                           & $5$ & $<12.79$& $<0.647$& $<0.608$\\
    \hline
    $M_{3}$          & $2$ & $>2.97$& $>6.26$& $>6.36$\\
    $(\mathrm{TeV})$ & $3$ & $>2.82$& $>5.95$& $>6.04$\\
    $ $              & $5$ & $>2.64$& $>5.57$& $>5.66$\\
    \hline
\end{tabular}
\end{center}
\caption{\label{tab:optimistic}
Same as Table~\ref{tab:conservative} but for the ``optimistic'' case.} 
\end{table}
The expected constraints on $f_i$ and $M_i$ in both the ``conservative'' and ``optimistic'' cases~\cite{AlAli:2021let} are calculated, and listed in Tables~\ref{tab:conservative} and \ref{tab:optimistic}.
Taking the case of $\sqrt{s}=3\;{\rm TeV}$ as an example, in this paper, the $2\sigma$ constraint on $M_0$ is about $27\%$ of $\sqrt{s}$ which is smaller than $\sqrt{s}$.
From the point of view of partial wave unitarity bounds, there is no sign that the validity of EFT has been violated.
However, if we assume that $f_i=c_i/\Lambda ^4$ and assume that $\Lambda \geq \sqrt{s}$, then $c_i\sim \mathcal{O}(4\pi)$~(for example $\left|f_0/\Lambda^4\right|s^2<12.1$ for $\mathcal{S}_{stat}=2$ and $\sqrt{s}=3\;{\rm TeV}$).
Thus the constraints are still in the strongly coupled scenarios, and one can expect that combined constraints of multiple processes or higher luminosities can further tighten the constraints.
Compared with the HL-LHC, where the combined~(combined of $pp\to \gamma\gamma$, $pp\to \ell^+\ell^-\gamma$, $pp\to \nu\bar{\nu}\gamma$ and $pp\to q\bar{q}\gamma$ channels) constraint on $M_0$ is about $22\%$ of $\sqrt{s}$~\cite{Ellis:2021dfa}, our result indicates that the muon collider is sensitive to gQGCs.

The uncertainties in this paper come from different aspects.
In MC simulation, higher order contributions are not included.
The beam induced background is also important, which typically leads to particles tangent to the trajectories, and is not yet included in \verb"Delphes" at this stage, therefore difficult to be considered.
Another uncertainty comes from the stochastic nature of training, e.g., randomly dividing the data into training, validation and test sets, random network initialization, random selection of a portion of the training data when calculating the gradient during training.
To study the projected sensitivities, one operator is considered at a time.
There are possible contributions from other high dimensional contributions.
Note that if the jets were from quarks, there will be possible interference between the SM and NP, and the interference of a dimension-12 operator is at the same order of gQGCs assuming the Wilson coefficients are at the same order.

\subsection{\label{sec:48}Compare AEAD with traditional method}

\begin{figure}[htbp]
\centering
\includegraphics[width=0.4\hsize]{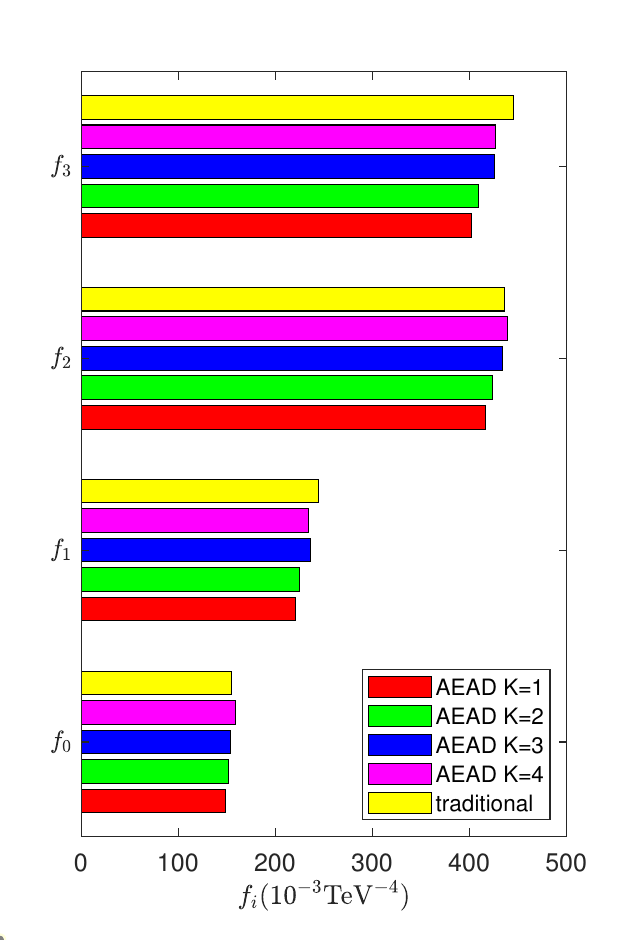}
\quad
\includegraphics[width=0.4\hsize]{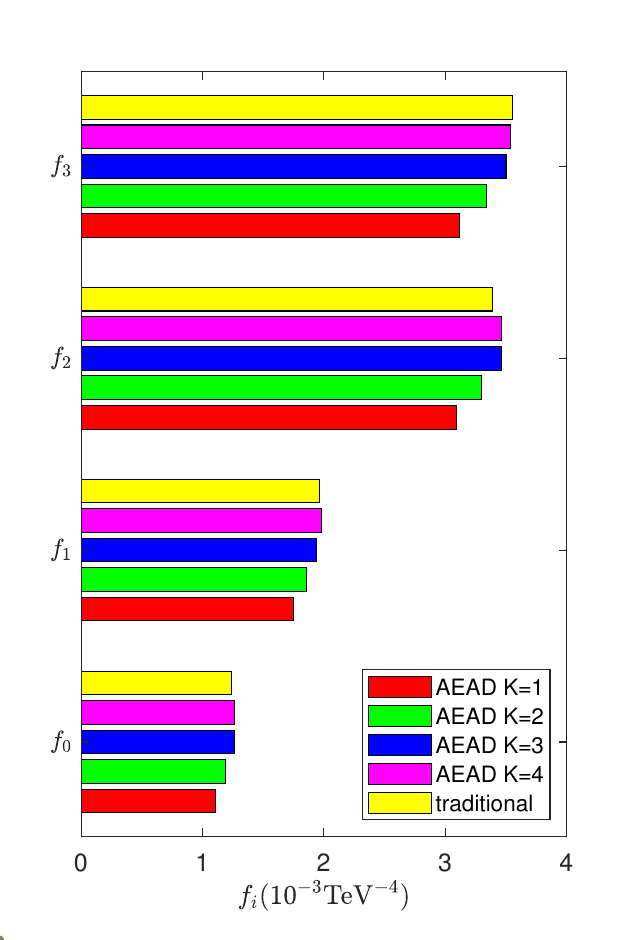}
\quad
\includegraphics[width=0.4\hsize]{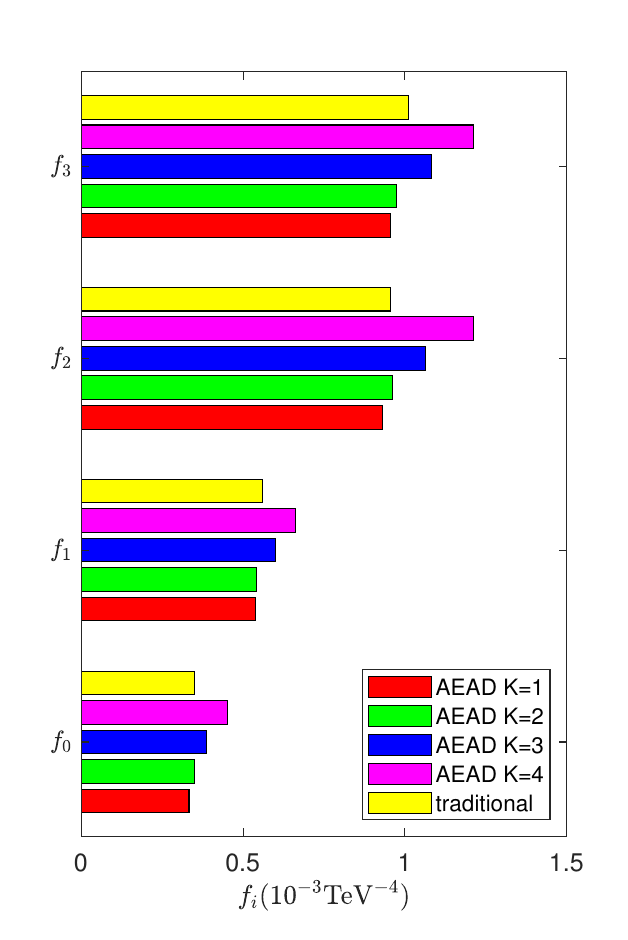}
\quad
\includegraphics[width=0.4\hsize]{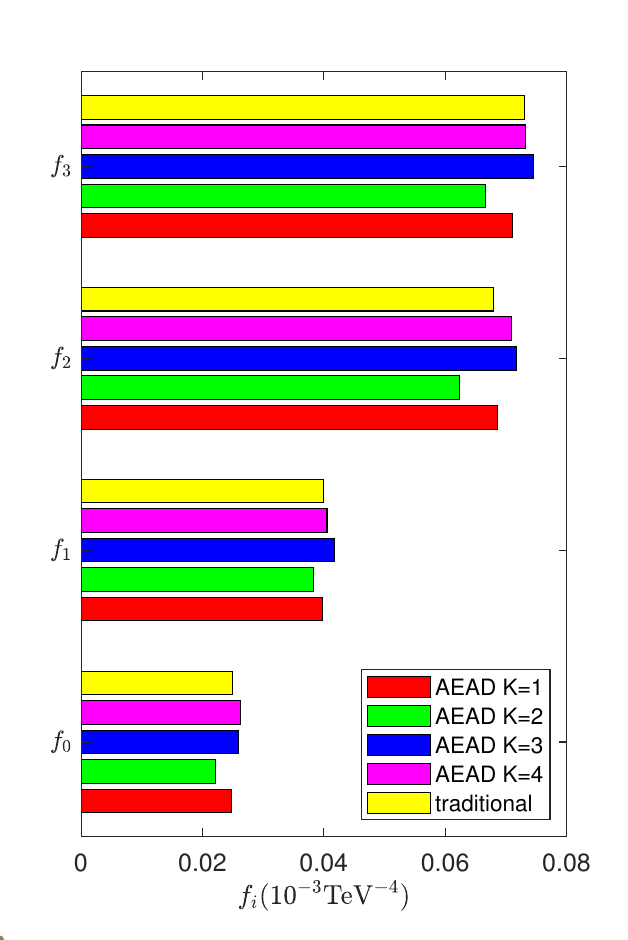}
\caption{\label{fig:compare}
The comparison of the expected constraints at $95\%$ C.L. level obtained by a traditional event selection strategy~\cite{Yang:2023gos}, and the AEAD event selection strategy.
The top-left panel corresponds to $\sqrt{s} = 3\;\rm TeV$, the top-right panel corresponds to $10\;\rm TeV $, the bottom-left panel corresponds to $14\;\rm TeV $, and the bottom-right panel corresponds to $30\;\rm TeV $.} 
\end{figure}

\begin{table}[htbp]
\centering
\begin{tabular}{c|c|c}
\hline
 $\sqrt{s}$ & $\slashed{p}_T$ & $m_{jj}$  \\
 $({\rm TeV})$ &  &  \\
\hline
 $3$   &  $>50\;{\rm GeV}$ & $>1\;{\rm TeV}$ \\
 $10$  &  $>100\;{\rm GeV}$ & $>3\;{\rm TeV}$ \\
 $14$  &  $>100\;{\rm GeV}$ & $>5\;{\rm TeV}$ \\
 $30$  &  $>200\;{\rm GeV}$ & $>10\;{\rm TeV}$ \\
\hline
\end{tabular}
\caption{The traditional event selection strategies at different energies~\cite{Yang:2023gos}. 
$\slashed{p}_T$ is the transverse missing momentum and $m_{jj}$ is the invariant mass of two hardest jets.
It is also required that $N_j\geq 2$ to calculate $m_{jj}$}
\label{table:traditionalcuts}
\end{table}

A comparison is made between the AEAD and the traditional method.
It can be seen that, the best expected constraints derived by the AEAD are generally better than those derived using the traditional method used in Ref.~\cite{Yang:2023gos}, which is listed in Table~\ref{table:traditionalcuts}.
The expected constraints on $f_i$ at $95\%$ C.L. level~($S_{stat}=2$) are compared with those from the traditional method~\cite{Yang:2023gos} in Fig.~\ref{fig:compare}, it can be shown that in all cases, the AEAD performs better.
It can be concluded that the AEAD can archive better results.
In particular, the AEAD always works better for $O_{gT,1,3}$ than $O_{gT,0,2}$, and the case of $\sqrt{s}=14\;{\rm TeV}$ is the worst case.

It is also noted that, only marginal gains are achieved by the AEAD with respect to the `traditional' event selection strategy.
However, compared with a traditional event selection strategy, AEAD is an AD algorithm.
It does not need the information of NP, in the training of the AE, only the information of the SM is used.
Although the AEAD does not utilize the information of NP, because of this it brings advantages.
AEAD can find signals of NP without knowing what NP signals to look for.
Except for that, AEAD simply find out the anomalous signals. 
Even if there was no NP, anomalous signals are noteworthy, they could be rare processes in the SM, or possible errors.

Another reason is that the final state of the process in this paper is simple. 
The ANN is good at finding potential patterns from a large and complicated feature space. 
It is expected that as the precision tests of the SM get more intensive, rarer processes with more final state particles will be considered, it will be more difficult for a traditional method to reveal NP signals from analysis of a large amount of possible observables, which is a better case to utilize the advantages of ANNs. 

\section{\label{sec:5}Summary}

With the potential for acceleration using quantum computing, the role of AE in searching for signals of NP becomes important, especially since NP has yet to show clear signs, and the search for NP cannot avoid dealing with increasing amounts of data for the foreseeable future. 
In this paper, the process of searching for NP signals using AEAD is proposed.
The procedure is independent of the content of the NP to be searched since only the SM data set is used in the training phase.

As an example, the process $\mu^+\mu^-\to \nu\bar{\nu}jj$ at muon colliders is considered, which is sensitive to the dimension-8 operators contributing to gQGCs.
The event selection strategy based on AEAD is studied, and the expected constraints on the operator coefficients are calculated.
It can be shown that, the constraints are generally tighter than those obtained by using a traditional event selection strategy.
Therefore, it can be concluded that the AEAD is effective in the phenomenological study of the SMEFT.
It is expected that the AE algorithm accelerated by quantum computers can be even more efficient in the future.

\begin{acknowledgements}
This work was supported in part by the National Natural Science Foundation of China under Grants No. 12147214, and the Natural Science Foundation of the Liaoning Scientific Committee No.~LJKZ0978.
\end{acknowledgements}

\bibliography{gqgcae}

\end{document}